\documentclass[aip,amsmath,amssymb,reprint,]{revtex4-1}

\usepackage{graphicx}
\usepackage{dcolumn}
\usepackage{bm}
\usepackage[utf8]{inputenc}
\usepackage[T1]{fontenc}
\usepackage{mathptmx}
\usepackage{etoolbox}

\usepackage{xcolor}
\usepackage{hyperref}
\usepackage{titlesec}
\usepackage{makecell}
\usepackage{gensymb}
\usepackage{subcaption}
\usepackage[normalem]{ulem}
\usepackage[english]{babel}
\def\selectlanguage#1{\relax}

\newcommand{\new}[1]{{\color{black}#1}}  
\makeatletter
\def\@email#1#2{%
 \endgroup
 \patchcmd{\titleblock@produce}
  {\frontmatter@RRAPformat}
  {\frontmatter@RRAPformat{\produce@RRAP{*#1\href{mailto:#2}{#2}}}\frontmatter@RRAPformat}
  {}{}
}%
\makeatother
\begin{document}

\preprint{AIP/123-QED}

\title{Low-dimensional platforms for single photon detection}
\author{Pushkar Dasika}
 
\author{Liza Jain}

\author{Varun Srivatsav Kondapally}

\author{Md Arif Ali}
\affiliation{Equal contribution}
\affiliation{ 
Department of Electrical Communication Engineering, Indian Institute of Science, Bangalore 560012, India
}%

\author{Medha Dandu}
\affiliation{%
Chemistry and Physics of Materials Unit, Jawaharlal Nehru Centre for Advanced Scientific Research, Bangalore 560064, India
}%
\affiliation{%
School of Advanced Materials, Jawaharlal Nehru Centre for Advanced Scientific Research, Bangalore 560064, India}
\author{Kausik Majumdar\textsuperscript{2,*}}
\email{kausikm@iisc.ac.in}


\begin{abstract}
A Single-Photon Detector (SPD) can detect extremely low intensity of electromagnetic wave - down to a single photon. Driven by the rapid developments in quantum information science and an increasing demand for ultra-low-light sensing across various domains, there is a need for transformative advancements in the design and development of SPDs. In this context, low-dimensional platforms, including quantum dots, superconducting nanowires and layered materials have emerged as crucial frontiers of research. This review explores the state-of-the-art of different low-dimensional SPD platforms, focusing on the engineering physics across their device architectures, performance parameters and application potential. By critically comparing the performance and addressing current challenges inherent to each low-dimensional platform, the review aims to outline future research directions to advance next-generation SPD technologies. 
\end{abstract}

\maketitle

\section{Introduction}
In an attempt to derive an expression for the spectral radiance of a blackbody in thermal equilibrium, Max Planck \cite{https://doi.org/10.1002/andp.19013090310} made the ad hoc assumption that the energy of an electromagnetic (EM) cavity is quantized. Albert Einstein in 1905 explained the photoelectric effect \cite{einstein1905erzeugung} using the concept of "light-quanta" that won him the Nobel prize in 1921. This interpretation marked a turning point in the history of physics, encouraging physicists to revisit the particle nature of light. A host of phenomena, such as reflection, refraction, interference, and diffraction, can be explained using the "wave-picture" of light. In contrast, phenomena like the photoelectric effect, the photovoltaic effect, and Compton scattering require a particle interpretation. The development of quantum mechanics in subsequent years resolved this discrepancy by proposing the concept of wave-particle duality, which states that light exhibits both wave-like and particle-like properties. The fundamental quantity in quantum mechanics is the wavefunction that elegantly combines the particle and wave nature. The eigenenergies of an electromagnetic cavity are the integer multiples (plus vacuum energy) of $h\nu$, called photons, where $\nu$ is the frequency of the radiation. In Dirac's bra-ket notation, an n-photon wavefunction is represented as $|n\rangle$. These are the eigenfunctions of an electromagnetic cavity. Naturally, the light we observe in our everyday world is a superposition of a very large number of eigenkets (Fock states: $|n\rangle$) with amplitudes that define the type of radiation, such as thermal radiation or laser radiation. Generating, manipulating, and detecting single photons (i.e., $|1\rangle$) is not trivial and is of huge technological relevance in the current era of quantum technologies. 

Since the 1980s, significant advancements have been made in understanding how quantum mechanics can be applied to solve practical problems, such as developing a computer that operates based on quantum principles \cite{hadfield2009single}. In recent years, efforts have been made to create quantum information and computational platforms that are compatible with digital and telecommunications technologies.  

Single-photon detectors (SPDs) are vital tools across cutting-edge fields due to their extreme sensitivity and picosecond timing resolution, which enable them to register the arrival time of individual photons. In quantum technologies, they are foundational, enabling secure communication through Quantum Key Distribution (QKD) and serving as the primary readout mechanism for photonic quantum circuits. Industrially, SPDs are transforming sensing and ranging through advanced single-photon LIDAR systems, offering extended range and high-resolution 3D mapping essential for autonomous vehicles and atmospheric monitoring. In life sciences, precision is indispensable for high-resolution imaging techniques such as Fluorescence Lifetime Imaging Microscopy (FLIM) and single-molecule detection, providing crucial insights into biological processes at the molecular level.

Given the need for SPD across various technologies, significant advancements have been made in recent years to create a viable technology for SPD. This is further fueled by the advancement of novel low-dimensional material systems.  In this review, we focus on the recent advances and challenges in various low-dimensional (0D, 1D, and 2D) SPD platforms. While silicon-based single-photon avalanche diodes (SPADs) have already achieved maturity and are widely available commercially, we do not cover SPADs as such here; instead, we use them as a benchmark for performance.

The rest of the paper is organized as follows: In section \ref{sec:pp}, we discuss the generic performance metrics of SPDs and their typical trade-offs. Next, we divide the available low-dimensional SPD platforms into three different classes: (1) semiconducting quantum well, nanowire, and quantum dot-based SPDs, (2) layered material-based SPDs, and (3) superconducting device-based SPDs. The superconducting SPDs are further divided into several sub-classes, such as (a) superconducting nanowire single photon detector (SNSPD), (b) transition edge sensor (TES), and (c) kinetic inductance detector (KID). A review of the mechanisms, performance, and challenges associated with each of these platforms is presented in sections \ref{sec:qd} to \ref{sec:sc}. We perform a benchmarking analysis of the performance metrics for these SPDs in section \ref{sec:bm}. Section \ref{sec:app} is devoted to different applications of SPDs, followed by a conclusion in section \ref{sec:con}.

\section{Performance Parameters of Single Photon Detectors and Trade-offs}\label{sec:pp}
The performance of a single photon detector is governed by a set of interrelated parameters that collectively define its sensitivity, speed, and noise characteristics\cite{hadfield2009single,eisaman2011invited,migdall2013single}. These figures of merit enable a useful comparison among various detector architectures and technologies, guiding the optimized design of devices for specific applications, such as quantum communication, quantum computing, low-light imaging, and time-resolved spectroscopy. The key performance parameters are defined below.
\subsubsection{Detection Efficiency}The detection efficiency ($\eta$) of a single photon detector represents how effectively it can convert an incident photon into a measurable electrical signal. \new{The internal detection efficiency (IDE) describes the probability that an absorbed photon leads to a measurable detection event, while system detection efficiency (SDE) correlates the measurable detection event to an incoming photon.} The overall $\eta$ is determined by several factors, including optical coupling, the extent of photon absorption, the internal quantum conversion, and the response of the readout electronics. For fiber-coupled detectors, the overall efficiency also includes transmission and coupling losses along the optical path. For free-space configurations, $\eta$ is often simplified to the product of absorption and internal quantum efficiency, along with the probability that the electrical pulse exceeds the detection threshold\cite{hadfield2009single,migdall2013single}. 

For an ideal detector, $\eta = 1$, i.e., all the incoming photons are successfully converted into sizeable electrical pulses. In practice, however, non-unity absorption, imperfect coupling, and electronic noise result in reduced detection efficiencies that vary across different technologies and wavelengths.
\subsubsection{Timing Latency and Rise Time}
The timing latency ($t_{latency}$) of a detector refers to the time interval between the photon's arrival at the detector and the instant when the resulting electrical pulse crosses a predefined threshold. This delay encompasses all the internal processes involved in converting the absorbed photon into an electrical signal, such as carrier generation and multiplication in semiconductor avalanche detectors, or quasiparticle diffusion and current redistribution in superconducting devices. The precise threshold level used to define latency is chosen based on specific requirements of the measurement system \cite{migdall2013single}.

The rise time ($\tau_{rise}$) characterises how fast the detector's electrical output increases once photon absorption occurs. It is typically quoted as the time taken for the electrical pulse amplitude to rise from $10\%$ to $90\%$ of its maximum value. $\tau_{rise}$ reflects the intrinsic speed of the detector as well as the bandwidth of the associated readout circuitry\cite{migdall2013single}.
\subsubsection{Timing Jitter}
Timing jitter represents the uncertainty in determining the exact arrival time of the detected photon. It quantifies the statistical variation in the delay between the photon's arrival and the generation of the corresponding electrical pulse, even when the conditions for arrival are identical \cite{hadfield2009single,migdall2013single}. Jitter is usually expressed as the temporal full width at half maximum (FWHM) of the instrument response function (IRF), though in some cases, the full width at $1\%$ of the maximum (FW$1\%$M) is also quoted to capture long tails in the timing response\cite{migdall2013single}.
Minimizing timing jitter is crucial for applications that require precise time resolution, such as quantum key distribution, fluorescence lifetime imaging, and time-correlated single-photon counting \cite{becker2005advanced}.
Low jitter can be achieved by enhancing material uniformity, carefully optimising the biasing conditions, and using high-bandwidth, low-noise readout electronics.
\subsubsection{Dead Time, Recovery Time, and Reset Time}
After a detection event, a single-photon detector undergoes a brief period during which it is unable to register subsequent counts. This time period, known as the dead time ($t_{dead}$), defines the maximum count rate that can be achieved by a detector. During dead time, the detection efficiency is zero \cite{hadfield2009single,migdall2013single}. The dead time may be caused by intrinsic processes or by external control systems.

Other closely related parameters are the reset time and recovery time. The reset time ($t_{reset}$) refers to the duration needed for the detector to return to its initial operating state after a detection event, while the recovery time ($t_{recovery}$) describes how quickly it regains full sensitivity to detect the next photon. These timescales determine the detector’s readiness for successive detections \cite{migdall2013single}.

Ideally, a single-photon detector would exhibit negligible dead time and an instantaneous reset, allowing for continuous photon counting. In practice, however, minimizing dead time often introduces trade-offs with other parameters such as elevated dark counts or incomplete recovery, which can result in false triggers or afterpulsing effects \cite{hadfield2009single,migdall2013single}.
\subsubsection{Dark Count Rate}
The dark count rate (DCR) is the average number of false detection events registered per second in the absence of any incident photons. These spurious detection events can originate from various processes such as thermal excitations, tunneling processes, blackbody radiation, or electronic noise from the readout circuitry. Reducing DCR typically involves operating at lower temperatures, optimizing material quality, and reducing the bias current. However, these strategies often come at the expense of reduced detection efficiency. Thus, achieving a balance between high detection efficiency and low DCR remains a critical challenge in the design and operation of single-photon detectors \cite{hadfield2009single,migdall2013single}.
\subsubsection{Afterpulsing Probability}
Afterpulsing refers to the occurrence of false detection events that follow genuine detection events, resulting from trapped charge carriers or incomplete recovery of the detector. For example, in a single photon avalanche detector (SPAD), when a photon is absorbed and an avalanche process occurs, a fraction of the carriers may become trapped in defect states within the detector material. These trapped carriers can get released after a delay and retrigger an avalanche, causing false counts correlated with the original event\cite{migdall2013single}. High afterpulsing probability increases noise and degrades temporal resolution. Minimizing afterpulsing involves optimizing quenching circuits and material engineering. Ensuring complete recovery of the detector before the next detection event further reduces the afterpulsing probability. However, this can prolong the dead time, and thus an inherent trade-off exists between afterpulsing probability and dead time\cite{migdall2013single}.
\subsubsection{Photon-Number-Resolving Capability}
Photon-number-resolving capability refers to a detector’s ability to distinguish how many photons are incident in a given detection event, rather than merely registering their presence or absence. While conventional single-photon detectors operate in a binary “click/no-click” mode, PNR detectors provide richer information about the photon statistics of light fields \cite{hadfield2009single,migdall2013single}.

PNR capability is particularly valuable in quantum optics, quantum communication, and photon-correlation studies \cite{becker2005advanced}, where accurate reconstruction of photon number distributions is essential. The performance of a PNR detector depends strongly on its detection efficiency, uniform response, and low crosstalk among detection channels. Multiplexing approaches—such as dividing the optical signal across spatial or temporal modes—can approximate PNR behavior by distributing photons among multiple single-photon detectors. However, these schemes may introduce additional dead time or complexity.

Ideal PNR detectors produce an output signal that is inherently proportional to the number of absorbed photons, as seen in technologies such as transition-edge sensors. Achieving accurate photon-number resolution thus requires balancing high detection efficiency, minimal noise, and stable response characteristics.

It is important to note that no single detector simultaneously optimizes all performance parameters; improvement in one aspect often comes at the expense of another. For instance, maximizing detection efficiency may increase the dark count rate, while improving photon count rate may require faster recovery and more complex thermal management. Thus, SPD design inherently involves a balance among sensitivity, speed, and other factors, depending on the specific application. This interdependence can be illustrated through two representative platforms. In layered materials-based single-photon detectors, mechanisms such as photogating (discussed in detail in Section \ref{sec:lm}) produce an output voltage spike whose strength increases with an increase in the time duration the photocarrier remains trapped. To reliably herald the detection of a single photon, one must allow the signal to pick up enough strength above the noise floor. However, such wait time prolongs the time until the detector resets to its original state. This creates a trade-off between the detection efficiency and the maximum photon count rate. In SNSPD, suppressing the system dark-count rate is usually accompanied by a reduction in the detection efficiency and a slight increase in the timing jitter. For example, introducing cold optical filters effectively blocks blackbody radiation from impinging on the detector but adds insertion loss to the optical signal, thereby reducing the system's detection efficiency. These filters also distort the signal pulse shape, often leading to an increase in the timing jitter \cite{shibata2015ultimate}. These examples demonstrate that detection efficiency, dark count rate, timing resolution, and reset time are inherently linked through the underlying device physics, with the precise trade-offs being platform-dependent.
\section{SPDs using \new{Semiconducting} Quantum well, Nanowire, and Quantum Dot based Devices}\label{sec:qd}

Quantum dots and nanowires are zero-dimensional and one-dimensional systems, respectively. Their extremely reduced dimensions in two (nanowires) or three (quantum dots) dimensions provide new avenues for interesting device control. Among many other applications, these materials have been explored for single-photon detectors ever since the boom in research towards low-dimensional materials and mesoscopic physics \cite{nanoNanotechnologyTimeline}. The reduced dimensionality provides efficient and confined electrostatics, which are used to modulate current flow through a current-conducting channel when photoexcited carriers are trapped in them. 
\subsection{Detection Mechanism and Existing Works}
Broadly, two classes of detectors have emerged in this category: devices that rely on current modulation in a field-effect device structure or a resonant tunneling diode structure.

\paragraph{\textbf{Field-Effect Device Based SPD:} \label{subsec:qdogfet}}

\begin{figure}
    \centering
    \includegraphics[width=\linewidth]{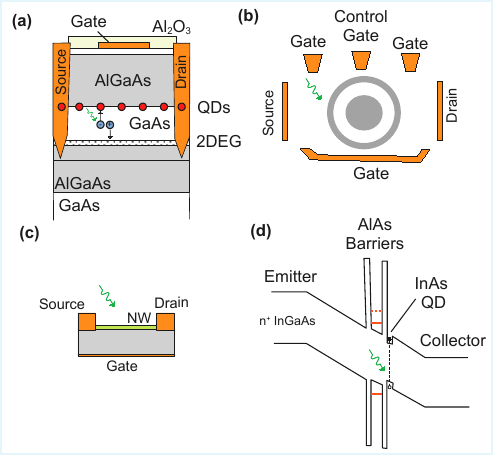}
    \caption{Various device structures to detect single photons using \new{semiconducting quantum dots and nanowires. The schematic diagrams of:
    (a) A quantum dot optically gated FET \cite{rowe2006single}. The quantum dots embedded between the gate and the 2D electron gas in the semiconducting layer modify the threshold voltage which is sensed to detect single photons. (b) A single electron transistor \cite{komiyama2000single}. The grey inner region represents the metallic inner core and the outer rings show the outer two Landau levels. Various gate area help to appropriately form 2D electron gas. (c) A semiconducting nanowire based transistor \cite{luo2018room}. Charge separation in the nanowir due to absorption of a single photon between the gate and the channel is used to detect single photons. (d) Resonant tunneling diode \cite{li2007quantum} where the photon absorbed by the quantum dot changes the electric field in the barrier region changing the resonant levels.}}
    \label{fig:quantumdot_devices}
\end{figure}

Quantum dot optically gated field-effect transistors (QDOGFETs) operate on the principle of photogating. 
In this structure, an active two-dimensional electron gas (2DEG) channel is separated from a layer of quantum dots with an active light-absorbing medium, as shown in Figure \ref{fig:quantumdot_devices}(a). 
An electric field is created in the absorbing layer by using a gate electrode. 
When light is absorbed by the light-absorbing medium, the electron-hole pairs are separated by the electric field.
The electrons (holes) join the 2DEG, while the holes (electrons) are trapped by the quantum dots.
The trapped holes in the quantum dots screen the gate field, bringing a change in the 2DEG channel current. 
When $N$ holes are trapped by the quantum dot layer, the change in current, $I_{DS}$, flowing in the 2DEG channel layer is given by \new{\cite{gansen2007photon}}

\begin{equation}
    \Delta I_{DS} = g_m\frac{eW}{\epsilon A}N \label{qdogfet_eq}
\end{equation}

where $g_m$ is the transconductance of the transistor, $W$ is the distance between the quantum dots and the gate, and $e$ is the fundamental electron charge. As can be seen from (\ref{qdogfet_eq}), the change in the measured current is directly proportional to the number of photons absorbed, which gives this type of single-photon detector photon number resolution capabilities\cite{gansen2007photon,kardynal2004low,kardynal2007photon}. Up to four photon number resolution has been reported \cite{gansen2007photon}.
Instead of using quantum dots, the intrinsic defects in the channel have also been used for single-photon detection \cite{kosaka2002photoconductance}. 
The usage of external quantum dots is preferred due to more device control that can be achieved.

A typical structure that has been used for such detection scheme is the GaAs/AlGaAs field-effect transistor, with a layer of InAs self-organised quantum dots \cite{rowe2006single,shields2000detection,gansen2007photon,kardynal2004low,kardynal2007photon}.
The major advantage of this device has been low dark count rates of about 0.003 counts per shot (ratio of number of counts of a particular current level in dark and the counts with illumination) \cite{rowe2006single}. A moderate to high internal quantum efficiency (IQE), in the range of 10\% \cite{kardynal2004low} to 68\% \cite{rowe2006single} has been reported. However, the external quantum efficiency (EQE) of the device has been limited to less than 2\% \cite{rowe2006single,shields2000detection}.  This is due to transmission losses at the gate electrode and absorption in the active layer \cite{rowe2006single,shields2000detection}. The speed of these devices has been generally reported to operate in the sub-Hz frequency range \cite{rowe2006single}. However, this speed is mostly limited by detection circuitry, and a higher speed of up to 400 KHz has been reported by using efficient circuitry to minimise the capacitance of the circuit lead, such as using a cryogenic radio frequency amplifier, which can be placed very close to the detector operating at low temperatures \cite{kardynal2004low,gansen2007photon}.

The main challenge in photon number counting is to achieve a uniform change in the channel current for every photon absorbed. 
The change in drain current can depend on the location of the quantum dot layer where the photon is absorbed, as well as external interference signals. To mitigate this problem, the area of the quantum dot has been minimised \cite{gansen2007photon} to make the electrostatics more uniform. However, doing this might affect the external quantum efficiency of the device.

Single-electron transistors have also been used to detect single photons\cite{komiyama2000single,sudha2025ultra,tabe2011single,nuryadi2006single}. For example, a silicon multidot FET has been used for single-photon detection\cite{nuryadi2006single,tabe2011single}. This device (shown in Figure \ref{fig:quantumdot_devices}(b)) has an ensemble of quantum dots connected through specific tunnel resistances and capacitances. Under light illumination, photon absorption can alter the number of electrons in the quantum dots, resulting in a change in device current. A sub-millimetre (up to 210 $\mu$m) wavelength single photon detector has been reported by placing a single-electron transistor in magnetic fields\cite{komiyama2000single}. When a sub-millimetre photon is absorbed by the quantum dot by cyclotron resonance, the excited electron-hole pair induces a strong polarisation within the dot. The intra-dot polarisation modulates the conductance of the single-electron transistor. These devices operate at mK temperature, with a quantum efficiency of less than 1\%, which has severely limited their expansion to more practical applications.

In addition to quantum dots, nanowires can also be used for single photon detection based on a similar principle \cite{luo2018room} of photogating as shown in Figure \ref{fig:quantumdot_devices}(\new{c}). Nanowires, due to their efficient gating, provide a good candidate for single-photon detection. CdS nanowire FETs have been reported \cite{luo2018room} to provide photon number resolution capabilities (up to 2 photons) at room temperature at UV frequency.
The detector offered 23 \% efficiency, although only a sub-Hz frequency could be achieved.

\paragraph{\textbf{Resonant tunneling-based devices (\new{RTDs}):} \label{subsec:rtd}}

In addition to devices which rely on a field effect device for amplification, a second class of devices used to detect single-photons based on modulation of tunnelling probability in a resonant tunnelling structure (shown in Figure \ref{fig:quantumdot_devices}(d)) has also been reported \cite{weng2015quantum,blakesley2005efficient,li2007quantum,li2008quantum}.
These devices operate by sensing a change in resonant tunnelling current through the device, which captures a photoexcited carrier from a quantum dot placed adjacent to the tunnel double barrier structure.

These devices have usually been reported with InAs quantum dots and AlAs/InGaAs/AlAs double barrier resonant tunnel diode operating at 4.2 K.
The wavelength of operation has been demonstrated from visible \cite{blakesley2005efficient} to telecommunications \cite{li2007quantum}.
The devices have achieved an overall efficiency of up to 2\%  with internal efficiencies of up to 15\%. 
Since resonant tunnelling devices are also based on the principle of electrostatic modulation of a barrier like a transistor, they can also show photon number resolution. Such a device has been reported\cite{weng2015quantum}, where up to 2 photons could be resolved at temperatures up to 77 K. However, the efficiency of the reported device is limited to 1\%.
\subsection{Challenges and Outlook}
The main challenge in these devices has been a low detection efficiency due to poor external quantum efficiency, although the internal quantum efficiency itself has been reported to be quite reasonable.
Device designs typically feature channels and quantum dots, which are obstructed by various device contacts, such as source/drain and gate terminals, resulting in the highest losses in external quantum efficiency. 
A more optimised optical design can potentially solve this problem.
Some efficiency is also lost in the recombination of the generated electron-hole pairs before they are collected in the device's channel.
Since the devices are based on charge trapping and detrapping, the frequency of operation and large timing jitter also pose major challenges. Furthermore, most devices are reported to operate in cryogenic temperatures, which limits their practical applications.
High-quality, scalable, and controlled fabrication of quantum dots and nanowires adds further limitations to this technology \cite{limame2024high}. 

\new{The research focus on this technology has been relatively less in the past decade or so, with more work focused on 2D materials and superconducting material-based detectors.} 
However, quantum dots, nanowires, and other transistor-based technologies still hold a good promise because of multiple reasons: (1) While most of the demonstrations use cryogenic temperature, in principle, these detectors are not inherently limited in their temperature of operation, unlike the superconducting materials, which need to operate below their critical temperature. (2) Advances in CMOS compatibility of III-V technology are expected to make it more suitable for large-scale production and integration in the already existing semiconductor technology deployments. (3) These detectors also hold the promise of photon number resolution capability.
\section{Single Photon Detectors using Layered Materials}\label{sec:lm}

Since the isolation of graphene in 2004 \cite{novoselov2004electric}, two-dimensional materials have transformed the landscape of condensed matter physics and optoelectronics \cite{liu2016van}. Their atomically thin nature, tunable bandgap, and strong light-matter interactions have enabled the exploration of exotic physics such as strong excitonic effects \cite{mueller2018exciton} and quantum confinement \cite{yang2017structural} that are unattainable in bulk semiconductors. Unlike bulk semiconductors bonded through strong ionic and covalent bonds, 2D materials comprise atomic layers held together by weak van der Waals forces in the out-of-plane direction. This allows for the mechanical exfoliation or "peeling" of individual layers \cite{novoselov2004electric}, which can be reassembled into heterostructures \cite{geim2013van} without the constraint of lattice matching, thus enabling the design of a wide range of interfaces with tunable band alignment, carrier dynamics, and optical responses \cite{novoselov20162d}.

Beyond their rich electronic tunability, layered materials can host correlated phases such as superconductivity \cite{xi2016ising} and charge density waves \cite{wilson1975charge}, making them suitable for exploring a range of rich nanoscale phenomena. These attributes have positioned layered materials as a versatile platform for single-photon detectors \cite{koppens2014photodetectors}, where the absorption of a single photon can trigger superconducting transitions, thermal fluctuations, or tunneling events within a few atomic layers. 2D material-based devices also offer opportunities for on-chip integration \cite{wang2024integration}. The following sections discuss the detection mechanisms employed in layered materials, summarize representative demonstrations, and highlight the challenges and opportunities in developing efficient single-photon detectors using these systems.

\begin{figure*}
    \centering

    \begin{subfigure}{0.49\textwidth}
        \centering
        \includegraphics[clip, trim=1.2cm 7.5cm 1.2cm 7.5cm, page=1, width=\linewidth]{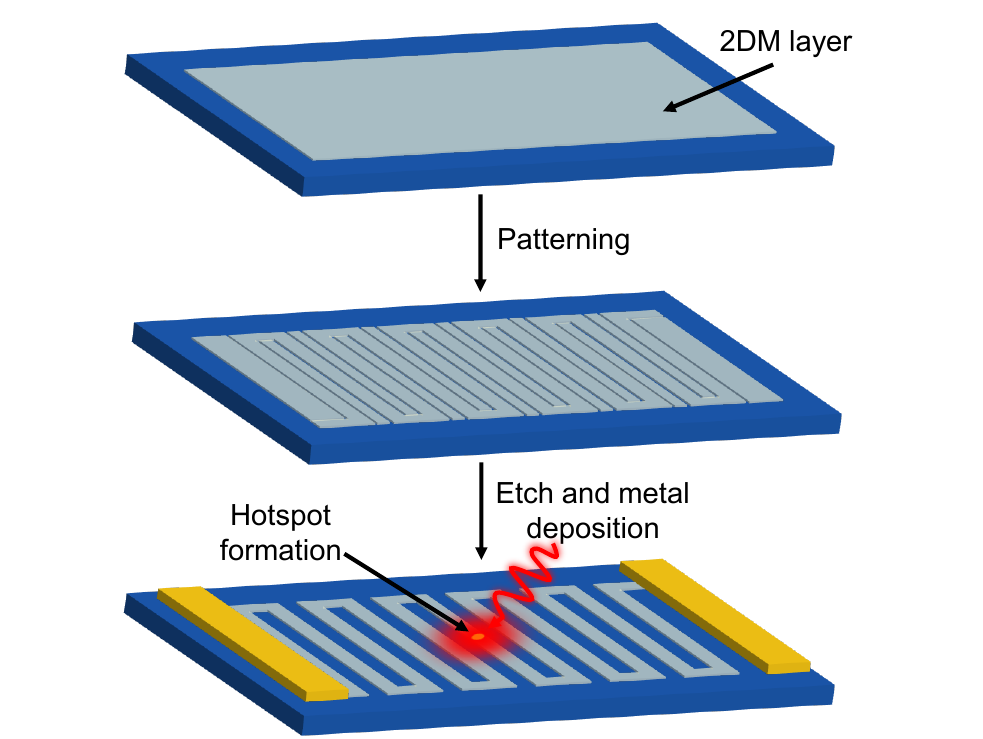}
        \caption{}
    \end{subfigure}
    \hfill
    \begin{subfigure}{0.49\textwidth}
        \centering
        \includegraphics[clip, trim=1.2cm 7.5cm 1.2cm 7.5cm, page=2, width=\linewidth]{Figures/layered_material_mechanisms.pdf}
        \caption{}
    \end{subfigure}
    \vspace{1em}
    \begin{subfigure}{0.49\textwidth}
        \centering
        \includegraphics[clip, trim=1.2cm 7.5cm 1.2cm 7.5cm, page=3, width=\linewidth]{Figures/layered_material_mechanisms.pdf}
        \caption{}
    \end{subfigure}
    \hfill
    \begin{subfigure}{0.49\textwidth}
        \centering
        \includegraphics[clip, trim=1.2cm 7.5cm 1.2cm 7.5cm, page=4, width=\linewidth]{Figures/layered_material_mechanisms.pdf}
        \caption{}
    \end{subfigure}

    \caption{A summary of detection mechanisms used in layered materials. (a) Photodetection using layered superconductor: A superconducting film is patterned and etched into a meander nanowire and contacted with metal pads. When a photon is absorbed, it locally breaks the Cooper pairs, forming a resistive hotspot that temporarily disrupts the superconducting path and results in a measurable voltage pulse \cite{metuh2025toward}. (b) Calorimetric detection mechanism: Incident photon raises the temperature of the active material, causing a change in resistance ($\Delta{R}$). The bottom panel displays the resistance versus temperature ($R$-$T$) curve, where the photo-induced temperature rise ($\Delta{T}$) results in a measurable change in resistance \cite{mckitterick2013performance}. (c) Schematic illustration of photodetection in a negative differential conductance (NDC) device. When the device is biased near the peak current, the photon absorption generates additional carriers, \new{perturbing the operating point and driving the device out of the NDC region. The purple arrows indicate the switching from the low-voltage branch to the high-voltage branch, producing a large, abrupt voltage change. The red arrows denote the subsequent relaxation of the device and its reset to the initial state under external bias conditions, enabling detection of subsequent photons.} \cite{nowakowski2025single}. (d) Photogating mechanism in a layered heterostructure. Photons are absorbed in the top absorber layer, generating electron-hole pairs. One type of carrier becomes trapped (localized) in a layer, causing a gating effect locally. The trapped charge modulates the conductivity of the underlying channel, resulting in an internal gain. The bottom panel shows the band diagram illustrating carrier trapping and channel modulation under drain bias. Dashed lines indicate energy bands after carrier trapping \cite{abraham2024room}}
    \label{fig:layered_material_detection_mechanisms}
\end{figure*}
\subsection{Detection Mechanisms}
\subsubsection{Superconducting detection}
In superconducting layered materials, the absorption of a photon perturbs the local superconducting order parameter by breaking Cooper pairs and generating excess quasiparticles\cite{tinkham2004introduction,semenov2001quantum}. This non-equilibrium excitation can locally suppress the superconducting gap, leading to a transient resistive state. While the basic principle parallels that of nanowire-based superconducting detectors, reduced dimensionality in van der Waals superconductors introduces distinct features such as enhanced thermal fluctuations, anisotropic energy relaxation, and electrostatic tunability of the critical temperature $\text{T}_c$ \cite{wang2023quantum}. These effects can modify both the sensitivity and recovery dynamics in comparison to conventional superconducting thin films. The fabrication of a meander nanowire using a layered material and the formation of a hotspot after photon absorption are shown in Figure \ref{fig:layered_material_detection_mechanisms}(a)
\subsubsection{Calorimetric detection}
In calorimetric photodetection, a photon is absorbed, resulting in a temperature change, which is then measured to determine the photon's energy \cite{mccammon2005thermal}. Unlike SNSPDs, no phase transition occurs, and thermal energy is measured directly. 

The heat balance equation is given as: 
\begin{equation}
    C_e\frac{dT_e}{dt} = P_{abs} - G_{e-ph}(T_e - T_{ph})
\end{equation}
where $C_e$ is the electron heat capacity, \new{$P_{abs}$ is the absorbed optical power,} $G_{e-ph}$ is the electron-phonon coupling, and $T_e$ and $T_{ph}$ are electron and phonon temperatures, respectively. When a photon of energy $E$ is absorbed, the electronic temperature rises by \cite{pretzl2020cryogenic}
\begin{equation}
    \Delta T_e = \frac{E}{C_e}
\end{equation}
A smaller $C_e$ leads to a larger $\Delta T_e$, which is important to have a sizeable signal. The system subsequently cools by electron-phonon coupling with a characteristic relaxation time $\tau_{th} = C_e/G_{e-ph}$ \cite{mather1982bolometer,mckitterick2013performance}.

The resulting temperature rise is converted into an electrical signal using different readout schemes. In bolometric-type devices, the resistance or Johnson noise of the absorber changes with temperature \cite{mckitterick2013performance} as shown in Figure \ref{fig:layered_material_detection_mechanisms}(b). In Josephson-junction calorimeters, the increase in $T_e$ suppresses the superconducting critical current $I_c(\new{T_e})$; when the device is biased near $I_c$, this suppression triggers a switching event, producing a voltage pulse corresponding to a single photon\cite{walsh2017graphene}.

The energy resolution of calorimetric detectors is limited by thermal fluctuation noise and scales as \cite{pretzl2020cryogenic}
\begin{equation}
    \Delta E_{FWHM} \approx 2.35 \sqrt{4k_BT^2C_e}
\end{equation}
Hence, a low heat capacity and weak electron-phonon coupling are crucial for achieving single-photon sensitivity \cite{mckitterick2013performance}. 
The main readout schemes are discussed as follows: 
\paragraph{Johnson noise readout}
In calorimetric SPDs, the absorption of a photon induces a transient rise in the electronic temperature ($T_e$) of the absorber, which reflects in its intrinsic voltage fluctuations, i.e., the Johnson or Nyquist noise. The mean-square value of voltage fluctuations across a resistor R in thermal equilibrium is given by \cite{fleischmann2020noise}:
\begin{equation}
    \langle V^2\rangle = 4k_BT_eR\Delta{f}
\end{equation}
where $k_B$ is the Boltzmann constant and $\Delta{f}$ is the measurement bandwidth. Upon photon absorption, the electron temperature $T_e$ rises, and this is reflected in the noise power spectral density $S_V = 4k_BT_eR$, which can be measured through a cryogenic low-noise amplifier chain.
\paragraph{Josephson Junction readout}
In this approach, the photon absorber, such as graphene, acts as a weak link in
a superconducting-normal-superconducting (SNS) Josephson junction \cite{walsh2017graphene}. The junction's critical current $I_c$ depends strongly on the electron temperature $T_e$. When an absorbed photon raises the electron temperature in the absorber, the calorimetric effect slightly reduces $I_c$. If the junction is biased near its critical current ($I_b \approx I_c$), this small change can trigger the junction to switch from the zero-voltage state to a resistive state, producing a measurable voltage pulse. The switching probability $\Gamma$ follows the following relation \cite{walsh2017graphene}:
\begin{equation}
    \Gamma \propto exp\left[-\frac{\Delta{U(I_b, T_e)}}{k_BT_{eff}}\right]
\end{equation}
where $\Delta{U}$ is the energy barrier for switching and $T_{eff}$ is the effective noise temperature. \new{$\Gamma$ determines how reliably the junction responds: a higher value increases the likelihood that photon absorption triggers a measurable voltage pulse, improving detection efficiency. The shape of the energy barrier $\Delta{U(I_b, T_e)}$ and $T_{eff}$ set the timing resolution: steeper barriers and lower noise provide faster, more consistent switching, while shallower barriers or higher $T_{eff}$ increase jitter. By carefully tuning these factors, SNS Josephson junction calorimeters can achieve both high efficiency and precise timing.}
\subsubsection{Negative Differential Conductivity-Based Detection}

Negative Differential Conductivity (NDC) occurs when the electrical current passing through a device decreases despite an increase in the applied voltage, \textit{i.e.}, $\frac{dI}{dV} < 0$. This nonlinearity can arise from resonant tunneling or moiré miniband effects in van der Waals heterostructures \cite{britnell2013resonant,leech2018theoretical}. 

The NDC phenomenon has been harnessed for photodetection, in which an absorbed photon perturbs the local carrier distribution and potential landscape, inducing bistable state switching between high- and low-conductivity branches \cite{nowakowski2025single}(Figure \ref{fig:layered_material_detection_mechanisms}(c)). This abrupt change in voltage provides a sizeable electrical response corresponding to individual detection events. Such detectors exploit the intrinsic instability of the NDC regime to achieve high responsivity without relying on superconductivity or cryogenic cooling; thus, the operating temperature can potentially be higher \cite{nowakowski2025single}. 
\subsubsection{Photogating mechanism}
In photogating, the absorbed photon generates electron-hole pairs; one carrier gets trapped in a defect state or an engineered trap, while the other carrier remains mobile \cite{fang2017photogating}. The trapped charge acts as a local gate, shifting the channel potential and producing a persistent current until it gets detrapped, yielding an internal gain\cite{fang2017photogating} as shown in Figure \ref{fig:layered_material_detection_mechanisms}(d).

This gain mechanism enables high responsivity and sensitivity to weak light. However, the long trap lifetimes that produce a large signal also introduce longer dead times \cite{fang2017photogating}. Despite this trade-off \new{with photogating},  layered material heterostructures \new{remain attractive for optimizing SPD design} due to their tunable interface trapping dynamics, strong carrier confinement, and compatibility with scalable fabrication processes \cite{liu2024photogating,roy2018number,abraham2024room}.

\subsection{Existing works}
Progress in single photon detection using layered materials followed a clear trajectory from theoretical proposals to proof-of-principle experimental demonstrations. Graphene has emerged as an essential material for single-photon calorimetry. The theoretical framework established by \textcite{mckitterick2013performance} demonstrated that graphene's ultra-low electronic heat capacity and weak electron-phonon coupling, originating from its vanishing density of states (DOS) near the charge neutrality point (CNP), enable highly sensitive cryogenic calorimetry via Johnson noise readout. Building on this theoretical foundation, \textcite{walsh2017graphene} proposed a superconducting-normal-superconducting (SNS) Josephson junction with graphene as a weak-link material. In their model, the absorbed photon raises the electron temperature in the graphene sheet, thereby suppressing the superconducting proximity effect and causing the Josephson junction to switch from a zero-voltage state to a resistive state. This concept was subsequently realized experimentally in a graphene-based Josephson-junction calorimeter, which achieved an intrinsic quantum efficiency of 87\% with dark counts less than 1 per second at operating temperature as high as 1.2 K, owing to the considerable electron-temperature rise ($\sim$2 K) enabled by graphene's ultra-low heat capacity\cite{huang2026thermal}. 

 Subsequent studies attempted further simplification in device architecture by exploiting superconductivity within the 2D material itself. When two graphene layers are stacked on top of each other with a relative twist angle of $1.1\degree$, known as the "magic angle", flat electronic bands are formed, giving rise to strong electronic correlations and gate-tunable superconductivity. \textcite{seifert2020magic} proposed using magic-angle twisted bilayer graphene (MATBG) directly as the superconducting material for calorimetric detection, eliminating the need for an external Josephson junction. Their model predicted broadband single-photon sensitivity with nanosecond response times and energy resolution better than 1 THz. In their 2024 work, \textcite{di2024infrared} demonstrated a proof-of-principle single-photon detector using superconducting MATBG. The device was illuminated with a highly attenuated 1550 nm laser at millikelvin temperatures and voltage-biased close to the critical current. Upon photon absorption, Cooper pairs in the MATBG were broken, driving a transient transition from the superconducting to the resistive state. To avoid permanent latching in the stable resistive state, the authors implemented a self-reset circuit that provided electrothermal feedback in which an increase in device resistance diverted current through a parallel load resistor, reducing Joule heating and allowing the device to cool back into the superconducting phase. The resulting voltage pulses were recorded using a low-noise amplifier, which demonstrated distinct photon-induced events and confirmed single-photon sensitivity.

 In a recent work, \textcite{nowakowski2025single} (2025) demonstrated single-photon detection via a novel mechanism exploiting negative differential conductivity (NDC) in moiré superlattices of bilayer graphene/hBN. By engineering a bistable conduction regime in the moiré minibands, the absorption of a single photon triggers a switch between high- and low-conductivity states, producing a detectable electrical signal. Since the operation relies on NDC rather than superconductivity, the device functions at elevated temperatures (up to $\approx$25 K) and across a broad spectral range from 675 nm to 11.3 \textmu m, demonstrating a promising route toward high-temperature, broadband single-photon detection in layered systems.

 Following the developments in graphene-based detectors, research attention has been extended to other layered materials such as NbSe$_2$. This material exhibits intrinsic superconductivity even in a few-layer form. In their 2025 work, \textcite{metuh2025toward} reported an experimental demonstration of few-layer NbSe$_2$ nanowires in a meandered geometry, exhibiting photon-induced resistive switching and a recovery time of ~135 ns, along with system timing jitter of nearly 1.1 ns. \textcite{zugliani2025single} reported the architecture with straight nanowire segments and hBN encapsulation, achieving dark-count rates below 1 Hz, timing jitter under 50 ps in selected traces, and clear single-photon detection at both 780 and 1550 nm. Together, these works mark significant progress toward scalable, high-performance layered-material SNSPDs.

 Additionally, van der Waals heterostructures have emerged as a versatile platform for single-photon detection, exploiting photogating effects to achieve high responsivity and low noise. In 2018, \textcite{roy2018number} reported a hybrid device combining bilayer graphene and MoS$_2$, in which the graphene channel was sensitized via a gapped bilayer and MoS$_2$ interface to achieve considerable optical gain ($\sim10^{10}$), an operating temperature of $\sim$80 K, a dark count rate as low as $\sim$0.07 Hz, and a linear dynamic range of $\sim$40 dB, although the photoresponse being relatively slow. More recently, in 2024, \textcite{abraham2024room} demonstrated a room-temperature SPD at the telecom wavelength (1550 nm) using a low-bandgap absorber ($\sim$350 meV) coupled to a sensitive 2D probe. They achieved an external quantum efficiency of $\sim$21.4\% (estimated to reach $\sim$42.8\% for polarised light) and a dark count rate of ~720 Hz at ambient temperature. These works illustrate the evolution of 2D heterostructure SPDs from low-temperature proof-of-concept devices to room-temperature telecom-band detection, underscoring how interface engineering and gain mechanisms can serve as an alternative to cryogenic superconducting architectures. 
\subsection{Challenges and Outlook}
Despite the remarkable progress made over the last decade, 2D-material–based single-photon detectors still face several challenges that hinder their transition from laboratory demonstrations to deployable quantum technologies. These issues span fundamental physical limitations to large-scale engineering barriers.

A fundamental limitation arises from the atomically thin nature of 2D materials. The intrinsic optical absorption per layer is extremely low, for instance, about 2.3\% for monolayer graphene \cite{nair2008fine}, which limits light-matter coupling and consequently the external quantum efficiency. This constraint is generic to most 2D systems, including transition-metal dichalcogenides (TMDCs), black phosphorus, and their heterostructures. Achieving strong light–matter coupling without compromising device speed or noise remains an active research frontier.

Material quality and interface control form another critical bottleneck. Due to the atomically thin nature of 2D materials, even trace impurities \cite{rhodes2019disorder}, wrinkles \cite{zhao2024wrinkle}, or trapped adsorbates \cite{beckmann2022role, haigh2012cross} at interfaces can dramatically alter local potential landscapes and introduce disorder \cite{hus2017spatially,rhodes2019disorder}. Such imperfections degrade carrier mobility \cite{wu2016defects}, increase noise, and lead to non-uniform gain across devices \cite{jiang2019defect}. In heterostructures, interlayer misalignment \cite{uri2020mapping} and contamination \cite{haigh2012cross} can significantly alter phase transitions \cite{nowakowski2025single} and carrier trapping mechanisms \cite{abraham2024room}, resulting in unpredictable photoresponse. Maintaining clean interfaces during exfoliation, transfer, and stacking is therefore essential but challenging to achieve with conventional polymer-assisted methods \cite{watson2021transfer}. Similarly, the lack of scalable synthesis routes for unstable materials, such as NbSe$_2$ or black phosphorus, limits reproducibility and reliability.

Scalability remains a persistent challenge when moving from proof-of-concept devices to large-area or pixel arrays \cite{zheng2025continue}. Wafer-level uniformity requires precise control over flake thickness, orientation, and contact geometry while maintaining both electronic and optical integrity \cite{xu2022growth}. Standard micro- and nano-fabrication processes, such as plasma etching, patterning, and deposition \cite{saifur2024effect, nakajima2025structural}, often introduce residues and cause surface damage that alters superconducting and transport properties. \new{These process-induced non-uniformities, besides intrinsic variations in contact resistances, complicate system-level integration for multi-pixel readout, leading to pixel-to-pixel variability in detector response and timing jitter \cite{anwar2025pixel}.} Moreover, cross-talk suppression in multi-pixel 2D SPD arrays requires careful electrostatic and thermal isolation \cite{allmaras2020demonstration}, which remains challenging to achieve without compromising compactness.

In calorimetric 2D SPDs, unwanted thermal leakage through substrate and contacts can drastically reduce sensitivity by allowing heat to escape before it can be registered as a measurable signal \cite{mckitterick2013performance}. \new{Thus, careful engineering of thermal coupling to substrates and metallic interconnects is required to avoid any degradation in detector sensitivity.}

Several operational constraints also limit practical adoption. Many 2D SPDs require cryogenic temperatures to function reliably. Maintaining such conditions adds cost and complexity, restricting applicability to specialized laboratory or space environments \cite{huang2026thermal,seifert2020magic,di2024infrared,nowakowski2025single}. Even in semiconducting or photogating architectures that can, in principle, operate at higher temperatures, managing thermal noise and ensuring long-term stability remain unresolved issues \cite{roy2018number,abraham2024room}. Environmental degradation caused by oxidation, moisture, or photo-induced chemistry further complicates field operation and storage.

Nonetheless, ongoing advances provide a path forward. Strategies that embed absorbers in micro/nano photonic structures \cite{li2019engineering}, such as microcavities \cite{furchi2012microcavity}, plasmonic structures \cite{muench2019waveguide}, photonic crystals \cite{noori2016photonic}, and metasurfaces \cite{guo2023hybrid}, have shown immense potential to overcome the limited absorption challenge. Directly integrating the detector with a waveguide or optical fiber can significantly enhance the light-matter interaction length, thereby helping to overcome the photoabsorption bottleneck in layered material-based SPDs. Balancing these approaches against unwanted increases in capacitance, noise, or fabrication complexity is an active area of research \cite{tao2021enhancing}. 

Encapsulation and passivation strategies using hexagonal boron nitride \cite{molaei2021comprehensive} and atomic-layer deposition \cite{sohn2024selective}, combined with vacuum packaging \cite{guo2023ultra}, can enhance stability against air and moisture. These developments, together with advances in cryo-CMOS readout integration and on-chip photonics, suggest that scalable and energy-efficient 2D-SPD platforms are within reach. \new{Achieving seamless integration requires careful and controlled impedance matching, low-noise electrical interfacing, and minimization of parasitic capacitances to preserve signal integrity and timing performances.}

At the device level, refined electrostatic, substitutional, \new{intercalation, and modulation} doping strategies \cite{zhao2017doping,li2022reducing,younas2023perspective} reduce contact resistance and enable tunable gain mechanisms. Controlled carrier multiplication through impact ionization \cite{li2024achieving} and photogating \cite{roy2018number,abraham2024room} offer routes to amplify weak photon-induced signals while minimizing readout noise. \new{However, in typical photogating architectures, the long-lived trap states responsible for high gain inherently limit response speed. From a defect-engineering perspective, passivating deep traps and introducing shallow or interfacial states can shorten carrier lifetimes while providing moderate gains. To this end, molecular decoration approaches demonstrated in layered materials such as MoS$_2$ and ReS$_2$ provide a promising route to selectively modulate trap states and suppress deep-trapping effects \cite{molina2016engineering,sun2018defect}. Device-level strategies such as channel-length scaling \cite{lee2020photoinduced}, vertical heterostructure engineering \cite{liu2024photogating, murali2019highly} for ultrafast carrier transit times by modulating transport mechanisms \cite{liu2024photogating}, band engineering \cite{murali2019highly}, and gate-assisted detrapping schemes\cite{park2025gate} for mitigating carrier lifetimes enable addressing the trade-off between responsivity and speed. However, these strategies may introduce some trade-offs in fabrication complexity, power consumption, and dark current. Active quenching circuits provide an additional system-level pathway to accelerate reset dynamics and improve temporal response. Further, forming moiré superlattices by twisting 2D materials} opens up the possibility of engineering band structures and correlated phases that enable broadband or high-temperature single-photon detection without relying exclusively on superconductivity \cite{nowakowski2025single}.

In summary, the current generation of 2D–material–based SPDs faces fundamental challenges related to limited absorption, interface disorder, operational constraints, and scalability issues during fabrication. However, the rapid evolution of materials processing, optical design, and heterostructure engineering provides an optimistic outlook. The combination of atomic-scale tunability, ultrafast carrier dynamics, and seamless compatibility with integrated photonics positions 2D materials as one of the most promising frontiers for next-generation quantum-light detection and imaging technologies.
\section{SPDs using Superconducting Devices}\label{sec:sc}
\subsection{Superconducting Nanowire Single Photon Detector (SNSPD)}\label{sec:snspd} 

A superconducting nanowire single photon detector (SNSPD) is an extremely sensitive device for sensing single photons. After the first demonstration of a superconducting wire strip utilized as a photon sensor by \textcite{gol2001picosecond} in 2001, SNSPDs have advanced in high system detection efficiency\cite{chang2021detecting,liu2024multispectral}, sensitivity from visible to mid-infrared wavelengths\cite{wollman2017uv,wang2019fast,wang2017design,verma2014high,wu2016nbn}, low dark count rate\cite{shibata2014quantum,shibata2016snspd}, low timing jitter\cite{vodolazov2019minimal} and fast detection time\cite{zhang201916}.

As in recent years, SNSPDs have become the gold standard for detecting single photons with high precision, enabling breakthroughs in quantum communication\cite{mao2018integrating,chen2020sending}, quantum computing\cite{couteau2023applications}, and advanced photonics\cite{gol2001picosecond,natarajan2012superconducting,chang2023nanowire}. Several superconducting materials, including NbN\cite{zhang2017nbn,dane2017bias,wu2016nbn,rosenberg2013high,yamashita2013low}, NbTiN\cite{zadeh2017single,miki2013high,schuck2013waveguide,tanner2010enhanced}, WSi\cite{verma2014high}, and MoSi\cite{reddy2019exceeding,liu2024multispectral,verma2015high}, have been utilized for fabrication of SNSPDs. These materials — with differing structures and intrinsic properties — have achieved system detection efficiencies (SDE) that exceed 90\% at the critical telecom wavelength of 1550 nm.
For example, \textcite{yang2017comparison} fabricated SNSPD based on NbTiN and NbN films on Si substrates and compared their properties. They concluded that although those SNSPDs had similar performances in terms of SDE and DCR, SNSPDs made of NbTiN showed better performance in terms of timing jitter and recovery time than those made of NbN.
Despite this impressive progress, a fundamental trade-off has been encountered between achieving unity intrinsic detection efficiency (IDE) and maintaining ultra-low timing jitter. This trade-off limits the simultaneous optimization of photon detection probability and timing accuracy, which are crucial for high-performance SNSPD applications.
\subsubsection{Detection Mechanism in SNSPD}
\paragraph{\textbf{Hotspot Model:}}
Currently, the detection mechanism is not fully understood. The first model is the generation of a hotspot by locally increasing the temperature. The energy of the photon is large enough to break the cooper-pair and generate a local tiny normal region.

\begin{figure}[h]
 \centering
\includegraphics[width=\linewidth]{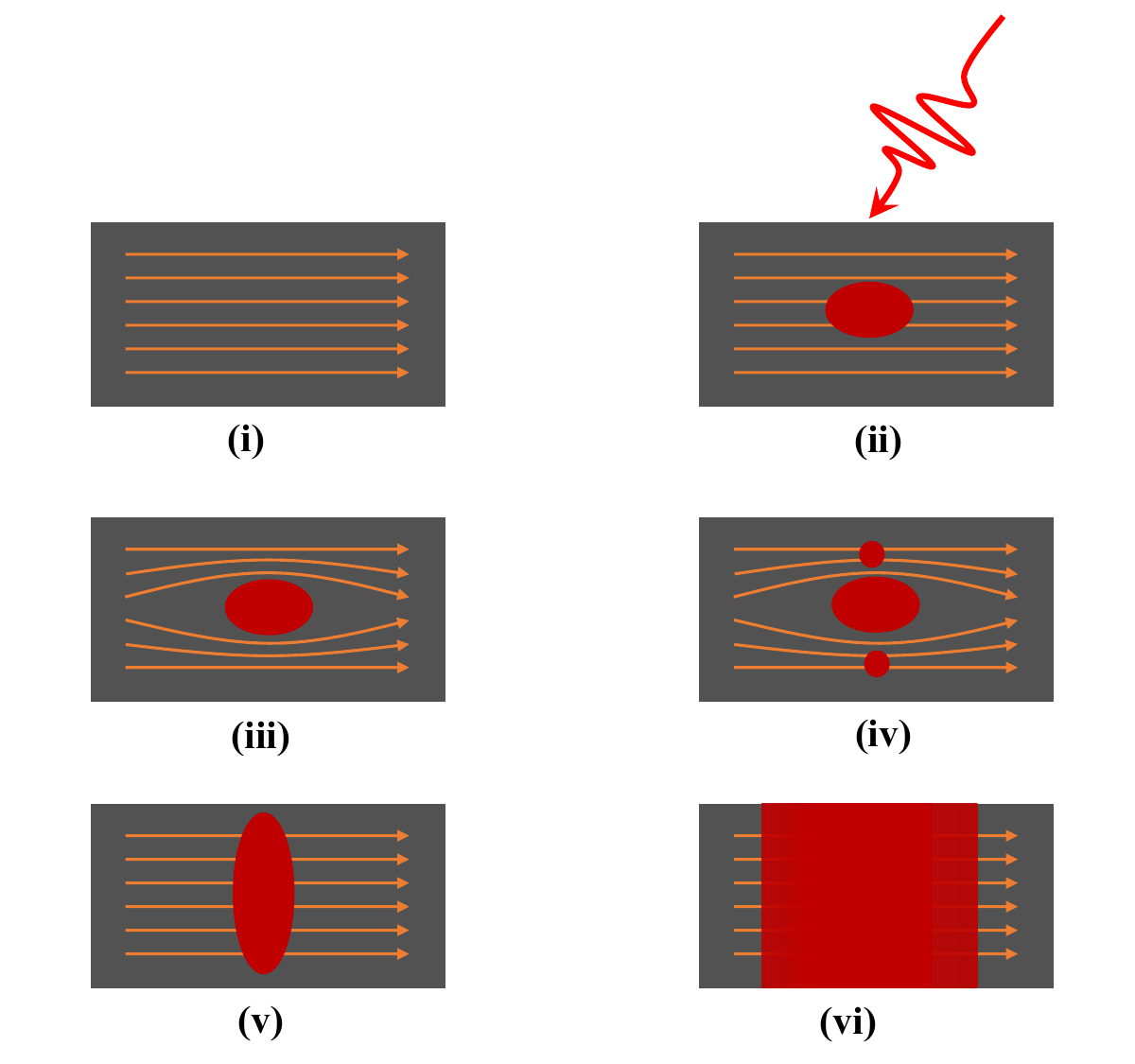}
\caption{The basic operation principle of the superconducting nanowire single-photon detector (SNSPD) indicating different stages of the hotspot model. \new{When a photon is absorbed by the nanowire, it breaks a large number of Cooper pairs, creating a localized nonequilibrium region—often referred to as a hotspot—within the superconductor. Although this hotspot does not initially extend across the full width of the nanowire, it perturbs the current distribution and forces the supercurrent to flow around the resistive region. As a result, the current density increases in the narrow side regions of the wire. If this redistributed current exceeds the local critical current density, superconductivity is suppressed across the entire width, leading to the formation of a resistive barrier.}}
\label{fig:SNSPD}
\end{figure}

The basic operation of SNSPD involves maintaining a thin superconducting nanowire at temperatures well below its critical temperature and biasing it with a direct current just under the  critical current of the nanowire. When a photon is absorbed by the nanowire, it disrupts many Cooper pairs, resulting in the formation of a localized hotspot in the superconductor. Although the hotspot is not large enough to span the entire nanowire width initially, it forces the supercurrent to bypass the resistive region. This diversion causes the current density in the side regions (sidewalks) of the nanowire to exceed the local critical current density, leading to the establishment of a resistive barrier across the wire, as shown in Figure \ref{fig:SNSPD}. The abrupt transition from zero to finite resistance generates a measurable voltage pulse across the nanowire, which serves as a readout signal indicating the detection of a photon. 

\begin{figure}[h]
 \centering
\includegraphics[width=\linewidth]{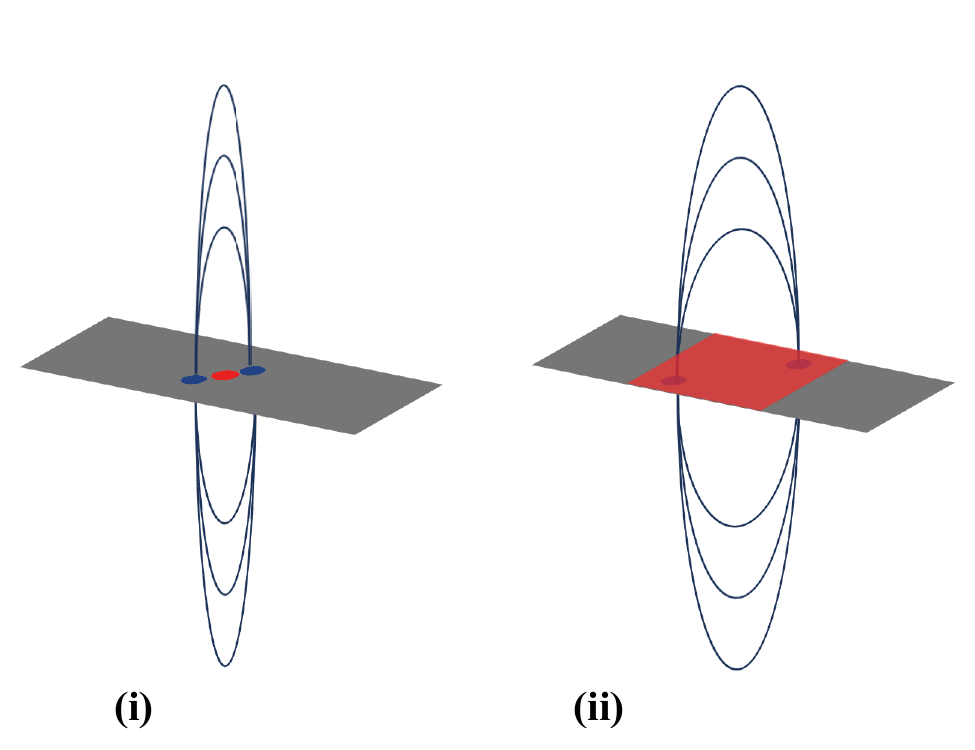}
\caption{The basic operation principle of the superconducting nanowire single-photon detector (SNSPD): vortex-antivortex model. \new{When a photon is absorbed, it first creates a vortex-antivortex pair. The current flowing through the wire then exerts a Lorentz force that drives the vortex and antivortex in opposite directions. This motion transitions the wire segment into the normal state, generating a measurable voltage pulse across the detector.}}
\label{fig:SNSPD_va}
\end{figure}

\paragraph{\textbf{Vortex-Antivortex Generation:}}

There are two suggested mechanisms for vortex-antivortex generation. The first one is the creation of a vortex-antivortex pair when the photon is absorbed. The current flow in the wire pushes the vortex and antivortex in the opposite direction by the Lorentz force. Then the wire transits into the normal state, creating a measurable voltage across the wire as shown in Figure \ref{fig:SNSPD_va}.

The second possible mechanism is the entry of a vortex inside the wire when the photon is absorbed, which surpasses the energy barrier for the vortex to penetrate the wire. Subsequently, the current helps to cross the wire throughout its entire width, hence generating a voltage pulse.
\subsubsection{Performance of SNSPD}
\paragraph{\textbf{System Detection Efficiency:}} 
Achieving close to 100\% SDE generally involves enhancing the nanowire material’s photon response capability, increasing photon absorption, and  improving the photon coupling into the active detector area.

For example, \textcite{marsili2013detecting} reported that a 4.5-nm-thick tungsten silicide (WSi) SNSPD thin film with a low superconducting energy gap exhibited improved intrinsic photon response capabilities, contributing SDE of 93$\%$. However, this thinness and smaller energy gap can increase the noise within the system, deteriorate the signal-to-noise ratio, and extend the recovery time of the device post-detection. Consequently, the timing jitter suffers, exceeding 70 ps in many reported devices. Similarly, a complex multilayer architecture that used a sandwich structure consisting of 6-nm-thick NbN layers separated by a 3-nm-thick SiO$_2$ layer was fabricated to boost photon absorption through optical constructive interference effects, thus improving system efficiency to 98$\%$ at 1590 nm and a system efficiency of over 95$\%$ in the wavelength range of 1530–1630 nm. Despite this advancement, the presence of the SiO$_2$ spacer altered current distribution and increased the kinetic inductance. This change, in turn, adversely affected the timing jitter by slowing down the detector’s reset and narrowing the bandwidth of the electrical pulse\cite{hu2020detecting}. Finally, the detector’s active area directly correlates with the efficiency of photon coupling from free-space optics or fibers. A large active area SNSPD based on 4.1-nm-thick MoSi with a 50 $\mu$m diameter was designed to improve photon coupling efficiency markedly from free-space or fiber optics\cite{reddy2020superconducting}. However, the expanded geometric size resulted in nonuniform current distribution and increased kinetic inductance, slowing the kinetic response and increasing timing uncertainty. As a result, these factors are restricting the detector’s timing resolution to about 70 picoseconds (ps)\cite{zhang2022nbn}.

\paragraph{\textbf{Dark Count Rate:}}
Reduction in DCR is  important for SNSPD performance for realizing application in quantum communication and sensing. A trade-off between sensitivity and low-noise is often unavoidable. The total system DCR comprises of two components: the intrinsic DCR and the blackbody-induced DCR (DCR$_{black}$). The system DCR is defined as the measured count rate when the input optical fiber to the cryocooler is blocked. The intrinsic DCR refers to the count rate observed when the fiber connected to the detector is physically disconnected inside the cryocooler, isolating the device from external optical input. In contrast, DCR$_{black}$ arises from room-temperature blackbody radiation that propagates through the optical fiber and reaches the detector. The system DCR is usually dominated by DCR$_{black}$ in the low bias region.

\textcite{shibata2016snspd} studied the effect of suppressing DCR using the various optical bandpass filters cooled at 3 K. They showed using a bulk commercial filter with a 10 nm bandwidth and transmission exceeding 85\% reduces the dark count rate by about 29 dB, at the cost of only a 2.4 dB drop in system detection efficiency. The DCR, primarily caused by background blackbody radiation entering through the optical fiber at room temperature, can be significantly reduced by incorporating a cold optical filter. \textcite{mueller2021free} showed that by filtering thermal photons, the DCR of a telecom-band SNSPD system can be lowered, without sacrificing efficiency or jitter. Here, a free-space–coupled SNSPD is demonstrated, offering high efficiency at 1550 nm, a dark count rate below 0.1 Hz, and timing jitter under 15 ps.

\paragraph{\textbf{Timing Jitter:}} Achieving low timing jitter is essential for applications requiring precise time-correlated single-photon counting, quantum key distribution, and high-speed optical communication. Two factors hinder the reduction of timing jitter. The first is electrical noise jitter, and the second factor is variable latency. Electrical noise in the readout electronics and amplifiers contributes to random fluctuations in the pulse timing signals. The electrical noise jitter can be minimized by designing detectors to operate at higher bias currents and with faster signal rise times, along with employing low-noise cryogenic RF amplifiers to reduce noise in the readout electronics. 

It was recently realized that fluctuations in the longitudinal position of the photon absorption lead to significant changes in the detection latency. The detection latency is defined as the time between the absorption of a photon and the registration of an electrical output pulse. This is due to the high kinetic inductance of typical nanowire structures, making the speed of RF pulse propagation along the wire to be a small fraction of the speed of light in vacuum\cite{korzh2020demonstration}.

In addition to these, electrothermal feedback has a substantial effect on timing jitter. It is the complex interplay between the line-edge roughness, localized heating,  cooling, and charge carrier dynamics during hot spot evolution that influences how fast and reliably the detector creates the output signal rise and returns to the superconducting state post-detection.

Despite these challenges, recent experimental efforts have pushed SNSPD timing jitter close to the physical limits under highly optimized conditions. A notable example is a short 5-$\mu$m-long NbN nanowire operated at 4 K with cryogenic, low-noise electronic readout, achieving system jitters as low as 2.6 ps at visible wavelengths and 4.3 ps at 1550 nm\cite{korzh2020demonstration}. This approaches the intrinsic physical timing limit constrained by kinetic inductance, electronic noise, and device recovery dynamics.

However, such devices are typically engineered with extremely small active areas to minimize kinetic inductance and reduce pulse propagation delays. Their small size and complex cooling and readout requirements preclude practical deployment in applications demanding high photon collection efficiency. Thus, scalability challenges remain a key barrier to translating these ultra-low jitter demonstrations into broadly useful systems.

\paragraph{\textbf{Reset Time:}}
The ability of SNSPDs to count at high rate — often cited as one of their most distinctive features — is generally hindered at high efficiency when using the standard readout scheme. This limitation arises due to a nonlinear interaction between the detector and the AC-coupled amplifier that reads out its signal, restricting high-rate performance. Additionally, reducing the active area to lower kinetic inductance enables faster reset times and higher count rates, but complicates efficient optical coupling. Similarly, increasing load impedance or using parallel nanowires to reduce inductance can introduce latching, thereby lowering detection efficiency\cite{rosenberg2013high}. Another factor is the thermal time constant that governs the intrinsic rate at which the nanowire cools and recovers superconductivity following photon absorption, thus establishing the fundamental lower limit for the SNSPD's reset time. The cooling of the resistive hotspot occurs when heat dissipates into the substrate and along the wire, allowing the temperature to drop back below the retrapping point\cite{zhang2018hotspot}. This cooling process is governed by the thermal time constant ($\tau_{th}$), which is established by the heat capacity and thermal conductance of the combined nanowire-substrate system\cite{yang2007modeling}.

Some research groups realized a high count rates reaching 1.5 Gcps\cite{craiciu_high-speed_2023, resta2023gigahertz, hao2024compact, zhang2024superconducting}. Recently, a 3-dB maximum count rate of 645 Mcps by a 64-pixel SNSPD array system was achieved by \textcite{fleming2025high}.

\subsubsection{Challenges and Outlook}
SNSPDs face several technical and practical challenges that limit their broader application space. Below we discuss three such key challenges:

\paragraph{\textbf{Mid IR Sensitive SNSPD:}}

SNSPDs have undergone extensive optimization for efficient performance across the ultraviolet, visible, and more recently, near-infrared (NIR) spectral regions. Extending their functionality into the mid-infrared (mid-IR) range, however, introduces new challenges due to the decreasing photon energy and the correspondingly lower sensitivity of superconducting materials at longer wavelengths.\cite{colangelo2022large}.

The implemented methods target two primary challenges in mid-IR photon detection: improving the sensitivity of superconducting thin films at longer wavelengths and addressing the limited fabrication yield for large-area devices. Attaining unity internal detection efficiency at these wavelengths generally requires jointly optimizing several interrelated parameters — such as minimizing the nanowire cross-section, employing materials with smaller superconducting energy gaps (and consequently lower critical temperatures), and reducing the number of quasiparticles necessary to trigger a hotspot. While thinner and narrower nanowires fabricated from low–critical temperature materials enhance detection sensitivity, they also significantly reduce the switching current—often below 1 $\mu$A—which degrades the signal-to-noise ratio and complicates the readout of voltage pulses that scale with the bias current\cite{patel2024improvements}.

\textcite{verma2021single} fabricated narrow nanowires of width 50 nm, from a high-resistivity 2.6 nm-thick WSi film, and achieved unity internal detection efficiency for wavelengths as long as 10 $\mu$m. This design approach simplifies the investigation of optimal device geometries and minimizes fabrication-related imperfections, allowing the detectors to achieve improved sensitivity to longer wavelengths. But to achieve a high system detection efficiency, the detector’s active area must also be scaled up. Overall, these factors underscore the delicate trade-offs involved in designing SNSPDs optimized for reliable and efficient photon detection in the mid-infrared region.

\paragraph{\textbf{Photon Number Resolution:}}

SNSPDs are highly attractive for their superior speed, low timing jitter, high efficiency, and low dark counts; however, their main limitation historically has been their limited photon number resolution. Standard SNSPDs are intrinsically binary, capable only of distinguishing between 0 and $\ge$ 1 photons, because the standard detection mechanism is intrinsically nonlinear, resulting in output pulses of nearly the same height regardless of whether one or multiple photons are absorbed\cite{los2024high, cheng2023100}. Photon-number-resolving capability is highly demanded for modern quantum applications, including photonic quantum computation, quantum communication, and the characterization of light sources.

To overcome their binary nature, various methods have been developed to transform SNSPDs into PNR devices, often classified as quasi-PNR detectors: a) Waveform Analysis (Intrinsic PNR): Instead of using arrays, conventional single SNSPDs can achieve PNR by analyzing the output signal waveform, which is encoded in the rising edge of the pulses\cite{los2024high}. b) Multiplexing: This approach uses multiple detectors\cite{cheng2023100}, where the sum of firing elements provides the photon number information. A large number of pixels is crucial for minimizing the probability of multiple photons firing the same pixel, thereby ensuring high PNR fidelity.

\new{Waveform analysis leverages the intrinsic response of a single SNSPD by analyzing pulse characteristics such as rise time, amplitude, and decay shape. Multiple photons create larger hotspots, producing measurably different waveforms that can be distinguished using principal component analysis. This method offers unambiguous one versus two photon resolution and partial discrimination up to five photons, while preserving the native SNSPD advantages of high speed and low complexity. 

The tradeoff between PNR detectors based on spatiotemporal multiplexing and those utilizing waveform analysis is primarily defined by a balance between resolvable photon capacity and timing precision. The hybrid spatiotemporal-multiplexing scheme, as demonstrated in the 100-pixel detector, achieves an unparalleled dynamic range by integrating an array of superconducting nanowires along a single waveguide, allowing it to resolve up to 100 photons\cite{cheng2023100}. In contrast, intrinsic PNR through waveform analysis, which relies on interpreting the rising edge of a single meander-shaped detector's output pulse, is currently limited to a much smaller range of up to 7 photons\cite{los2024high}. The limitation in the multiplexed approach stems from the statistical probability of multiple photons hitting the same pixel, which can cause "missing" counts at very high fluxes, whereas waveform analysis is limited by the jitter-to-risetime ratio, where arrival-time distributions of pulses for different photon numbers eventually overlap and become indistinguishable.}

\new{\paragraph{\textbf{Electrical Interfacing:}}

Electrical interfacing presents a significant challenge in SNSPD systems. These detectors require low-noise, high-bandwidth readout circuitry that operates at cryogenic temperatures, necessitating specialized impedance-matched microwave wiring and, in many cases, superconducting interconnects to suppress Johnson–Nyquist noise and minimize crosstalk. Despite ongoing progress, the understanding and optimization of SNSPD electrical behavior and readout architectures remain areas of active research, and existing design approaches are not yet fully mature\cite{ma2024research}.
A central challenge lies in converting the detector’s intrinsic electrical response into a usable signal. When a photon is absorbed, a transient resistive hotspot forms in the nanowire, momentarily diverting the bias current and generating a small voltage pulse. This weak, ultrafast signal must be amplified and conditioned without degrading its timing precision or signal-to-noise ratio, enabling reliable processing by conventional room-temperature electronics. In many advanced systems, the signal must also interface directly with photonic components located within the same cryogenic environment\cite{toso2021electronics}.


\paragraph{\textbf{Optical Coupling with Silicon Photonics:}}

Efficient optical coupling to SNSPD presents another challenge, largely because of the pronounced mode mismatch between the sub-micron SNSPD nanowire and silicon waveguides. This requires precise evanescent coupling or grating couplers, often resulting in coupling losses at every interface. Refractive index contrast between the cryogenic superconducting thin films (NbN, NbTiN) and silicon nitride/silicon waveguides demands careful multilayer design to avoid unwanted reflection. Achieving efficient coupling, therefore, requires carefully engineered, low-loss, broadband grating or edge couplers\cite{yang2023design, son2023highly}. Moreover, fabrication imperfections—such as sidewall roughness, thickness variations, or material non-uniformity can substantially reduce coupling efficiency. Ensuring high-efficiency optical power transfer between such dissimilar waveguide structures is essential for enabling low-loss interconnections and reliable integration across heterogeneous photonic platforms.}

\paragraph{\textbf{High-T$_\text{c}$ Superconductor Based SNSPD:}}
Recent experimental demonstrations using high-$T_c$ superconductors have shown that single-photon sensitivity can be extended far beyond operating temperatures of conventional SNSPDs. Two notable studies used cuprate-based superconductors. \textcite{merino2023two} used a two-dimensional cuprate superconductor $\text{Bi}_2\text{Sr}_2\text{CaCu}_2\text{O}_{8+x}$ (BSCCO) to realize a proof-of-concept single-photon detector under 1550 nm illumination operating at 20 K. The devices were fabricated as a few-layer BSCCO flake encapsulated with a top hBN and patterned into nanowires using a helium-focused ion beam (He-FIB). These non-optimized devices exhibit a slow reset time ($\sim$ ms) and a low detection efficiency ($\sim 10^{-4}$). Similarly, \textcite{charaev2023single} fabricated and characterized nanowires in BSCCO and $\text{La}_{1.55}\text{Sr}_{0.45}\text{CuO}_4\text{/La}_2\text{CuO}_4$ bilayer films, demonstrating single-photon response with BSCCO devices operating up to $\sim$ 25 K and LSCO devices near $\sim$ 8 K. Their studies highlight the importance of vortex/ fluctuation-assisted switching and show clear, bias-dependent photon-count statistics while leaving detailed metrics such as jitter and dark-count rate largely unreported.

Despite demonstrating single-photon sensitivity at telecom wavelengths and at operating temperatures far above those of conventional SNSPDs, cuprate-based detectors face several open challenges. Fabrication of ultrathin BSCCO/LSCO films is challenging due to the stringent control required for stoichiometry, mechanical stability, and interface cleanliness. Furthermore, the microscopic details of the detection mechanism are not well understood, which limits the development of single-photon detectors with deterministic and predictable characteristics \cite{merino2023two, charaev2023single}. Consequently, while current studies establish clear proof-of-concept for single-photon detection, cuprate detectors still require advancement in materials growth, device engineering, and mechanistic understanding before they can become reliable, high-performance single-photon technologies.

In addition to the above-mentioned challenges, there are several other aspects that need to be addressed in the near term. Scaling active area is one of the key challenges of SNSPDs. Creating larger SNSPDs for applications that require active areas ranging from hundreds of micrometers to centimeters necessitates highly uniform superconducting films, consistent detection performance, and the ability to integrate large-scale SNSPD arrays. Getting uniformity during fabrication and hence to avoid current crowding also remains as one of the fabrication challenges. Variations in film thickness, nano-fabrication defects, and meander bends can produce constrictions or current crowding, reducing critical current and uniformity, which decrease efficiency and reliability. Attaining near-unity detection efficiency, particularly for longer wavelengths, continues to be hindered by constraints in optical absorption and internal conversion efficiency within the nanowire. 

Addressing these challenges will open the door to new capabilities—including longer-wavelength photon detection, higher-temperature operation, and more accessible, cost-effective fabrication—ultimately making SNSPD technology more practical for quantum optics, quantum communication, quantum computation, and other advanced applications.

\subsection{Transition Edge Sensor (TES)}\label{sec:tes}
The TES is a low-temperature, highly sensitive thermometer that can detect very small temperature changes. At its core, there is a thin superconducting film whose temperature is maintained at the cusp of the transition. The sharp superconducting transition, accompanied by a low heat capacity, enables them to convert single-photon energies into detectable electrical pulses.

The basic design idea of the TES detector is to thermally couple a superconducting film to a photon-absorbing layer while weakly coupling it to a cryogenic bath. An example fabrication process is as follows: the process starts by taking a Si wafer and growing SiO$_2$ on it. Then, a layer of low thermal conductivity material, usually silicon nitride (SiN$_\text{x}$), is deposited on SiO$_2$. To achieve the desired $T_c$ ($\sim100$ mK), a superconducting bilayer (superconducting-normal) is deposited on SiN$_\text{x}$. Next, Nb leads are patterned on top of the film that serve as electrodes. Lastly, the film is capped with an absorber that is designed to absorb the desired wavelength\cite{ascheron_transition-edge_2005}.

\subsubsection{Detection Mechanism in TES}
The two crucial processes of the single-photon detection event are pulse generation and recovery. For successful pulse generation, the energy needs to be converted into an electrical signal that is subsequently amplified and read out. For efficient recovery, a robust mechanism is required to bring the detector back to the operating point. When a photon is incident on the TES detector, it is absorbed and the energy is transported to the superconducting layer as heat. This increases the film's temperature by creating a temporary hotspot, which drives the detector into the normal state. The increase in resistance of the film decreases the current through it, which is amplified and read out.

To efficiently recover the operating point of the detector, a negative feedback mechanism is used. To this end, the superconducting film is biased with a constant voltage, V$_{\text{bias}}$. This is usually done by placing a small shunt resistor in parallel to the detector and applying a constant current source to the circuit. The voltage bias is adjusted to stabilize the superconductor temperature on the transition edge. In equilibrium, the Joule heating by the biased superconducting film is dissipated by the heat absorption of the substrate. Now, when a single-photon absorption occurs, the heat increases the resistivity, which reduces the Joule heating $(\propto V^2/R)$. This is possible as the superconducting film is biased at a nearly constant V$_{\text{bias}}$. The heat dissipation through the substrate dominates the reduction in the Joule heating, returning the detector back to the quiescent point. This is known as Negative Electrothermal Feedback (NETF) \cite{irwin_application_1995,cabrera_detection_1998}.

\begin{figure}[h]
\includegraphics[width=1.7\linewidth]{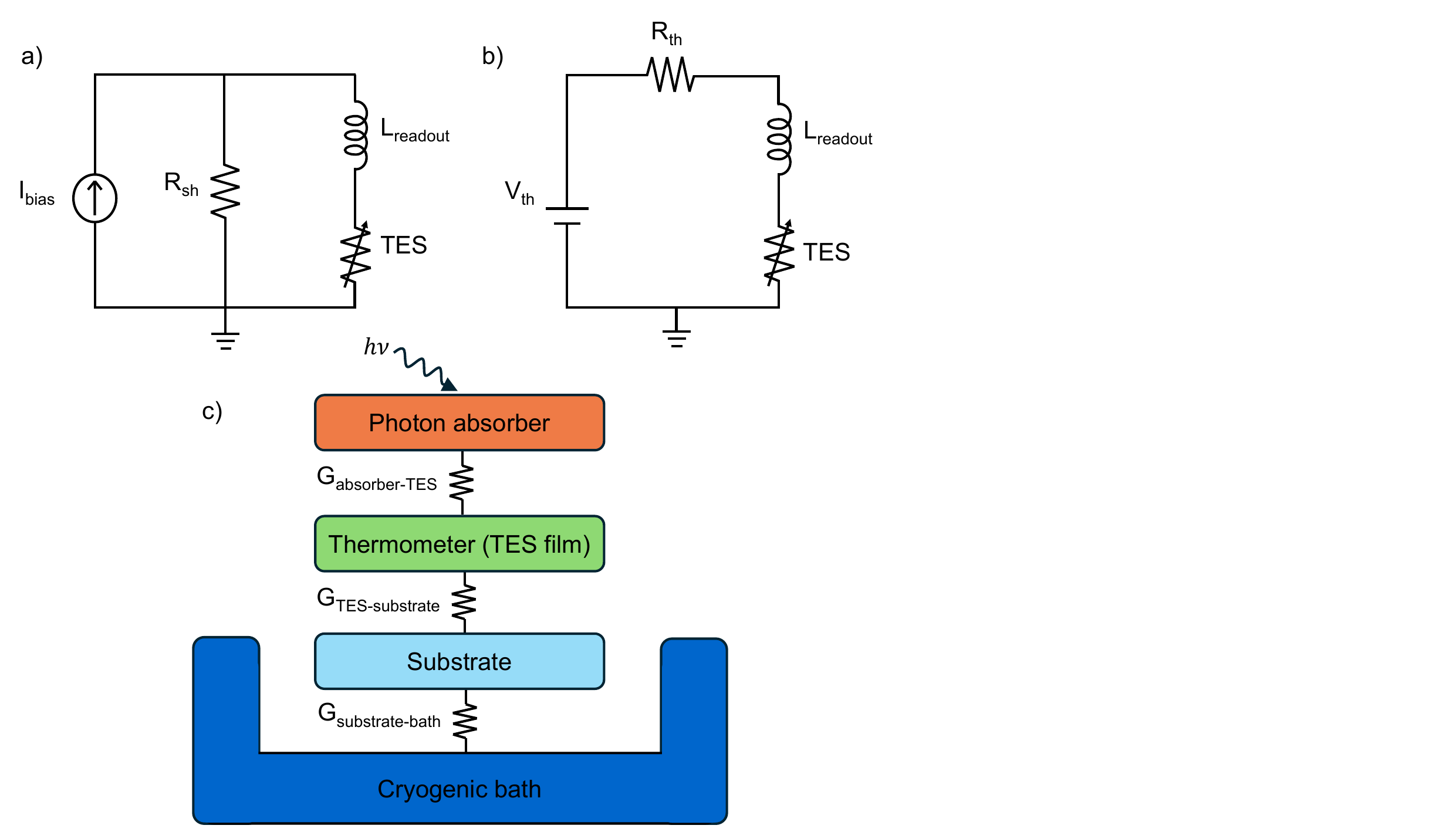}
\caption{\new{Electrical circuit and thermal model for transition edge sensors. a) A constant bias current $I_{bias}$ is applied across a shunt resistor $R_{sh}$ in parallel to the series combination of the readout inductance (SQUID inductance) $L_{readout}$ and the TES. b) The Thevenin equivalent of the electrical circuit in a). c) Thermal model of the TES detector. The incoming photon first encounters the absorber layer situated on the top of the TES stack and gets absorbed. The energy is transferred to the superconducting film connected to the absorber through a thermal conductance of $G_{absorber-TES}$. After generating an electrical spike, the film loses the heat to the substrate (through the contacts and the bottom substrate) through a thermal conductance of $G_{TES-substrate}$ and resets the detector. The substrate is directly coupled to the cryogenic bath through a thermal conductance of $G_{substrate-bath}.$}}
\label{fig:TES}
\end{figure}

 To read out and amplify the change in current from a single-photon event, an inductive coupling to a series of Superconducting Quantum Interference Devices (SQUIDs) is generally used. A comprehensive overview of the SQUID readout architecture can be found elsewhere \cite{ascheron_transition-edge_2005}. Schematic of the electrical circuit, its Thevenin equivalent, and the thermal model of the detector are shown in Figure \ref{fig:TES}. The process is mathematically described by two coupled differential equations \cite{ascheron_transition-edge_2005}:

 \begin{equation}
     P_{photon} = C\frac{dT}{dt}+P_{bath}-I^2R(T,I)
 \end{equation}
 \begin{equation}
     IR(T,I) = -L_{readout}\frac{dI}{dt}+V_{th}-IR_{th}
 \end{equation}
where C is the heat capacity of the superconducting film, T is the temperature, $P_{bath}$ is the dissipation of the power to the bath, I is the current through the detector, $V_{th}$ is the Thevenin equivalent voltage, $R_{th}$ is the Thevenin equivalent of the resistance, $L_{readout}$ is the coupling inductance to the SQUID amplifiers, and $R(T,I)$ is the detector resistance.

\subsubsection{Performance of TES}
\paragraph{\textbf{Detection Efficiency:}} 
Transition Edge Sensors have been demonstrated with near-unity internal quantum efficiencies \cite{fukuda_titanium-based_2011,lita_counting_2008,miller_demonstration_2003}. However, achieving near-perfect system detection efficiency (SDE) is non-trivial as losses can occur due to various factors such as misalignment of the photon path, coupling loss, and reflection from the detector surface. These challenges can be overcome by introducing some design modifications. For example, to reduce reflection from the detector, TES were fabricated between a high-reflectivity mirror and an anti-reflection layer \cite{rosenberg_performance_2005,lita_counting_2008,fukuda_titanium_2009}. Later, \textcite{fukuda_titanium-based_2011} demonstrated a 98\% SDE by resin (with refractive index matching the fiber core) embedding the optical fiber into an optical TES cavity for 850 nm photon wavelength. The above-mentioned techniques need meticulous manual alignment and face some practical challenges. To overcome these challenges, \textcite{miller_compact_2011} proposed a self-alignment technique and were able to show high SDEs for 805, 850, and 1550 nm wavelengths. Later, using the same technique, \textcite{kobayashi_development_2019} demonstrated an 83\% SDE at 940 nm wavelength based on Ti/Au bilayer. More recently, \textcite{li_development_2022} demonstrated a 55\% SDE at 1550 nm using this technique.

\paragraph{\textbf{Wavelength:}}
The detection of a specific wavelength or a band of wavelengths is enabled by using a particular absorber and an optimal design. Gamma-ray TES detectors use bulk Sn absorbers (which have low heat capacity in the superconducting state, maintaining good energy resolution) thermally coupled to the TES film \cite{bennett_high_2012}. The thickness of the Sn absorber is generally a few hundred microns to provide enough stopping power. Similarly, X-ray TES detectors use thick Bi or Au/Bi absorbers, owing to Bismuth's semi-metallic nature \cite{taralli_acdc_2020,brown_absorber_2008}. In contrast, the detection of UV, optical, and NIR wavelengths is achieved by embedding the TES film in an optical cavity. As discussed in the previous subsection, by optimizing the optical cavity for a specific wavelength, a near 100\% SDE can be achieved. However, the same design can perform poorly for a different wavelength due to effects such as cavity interference and other losses \cite{lita_counting_2008}. The reflected power as a function of incident wavelength shows a fringing pattern as expected from the constructive and destructive interference of photons in the TES cavity. This has obvious implications for SDE as the reflected power contributes to the system loss. Studies on the dependence of SDE on wavelength show how photon absorption efficiency varies with wavelength and reveal some unexplained low-energy peaks (false peaks) when excited with non-resonant wavelengths, attributing the origin to some other loss mechanisms \cite{hattori_optical_2018,hattori_optical_2019}. Far-infrared TES detectors use sophisticated designs to efficiently couple the incident radiation to the absorber and film \cite{mauskopf_development_2008,holland_scuba-2_2013}. It is necessary to lower the superconducting energy gap of the film (by the proximity effect) to enable long-wavelength detection.

\paragraph{\textbf{Dark Count Rate:}}
Optical TES detectors generally exhibit very low DCR, typically on the scale of 1 mHz \cite{xia_characterization_2025}. DCR is classified into two types based on its origin: Intrinsic to the detector and originating from external components. A study from 2015\cite{dreyling-eschweiler_characterization_2015} identified various types of events contributing to dark counts, such as environmental electromagnetic interference, cosmic rays, warm fiber end, etc, in an optical TES setup operating at 1064 nm. By performing pulse shape analysis, high-energy cosmic rays and electrical-noise events can be filtered out at the cost of a slight increase in the dead time fraction \cite{dreyling-eschweiler_characterization_2015}. Recently, \textcite{manenti_dark_2024} further analyzed the nature of different types of dark counts using singular-value decomposition, principal component analysis, and k-means clustering. More studies on the origin of background noise sources are essential to further improve the noise performance of the setup.

\paragraph{\textbf{Photon Number Resolution:}}
TES detectors are inherently photon number resolving, and numerous works have characterized their PNR capabilities \cite{lita_counting_2008,fukuda_titanium_2009,fukuda_titanium-based_2011,mitsuya_photon_2023,rosenberg_performance_2005,lita_superconducting_2010}. Extracting the number of photons out of the detected voltage pulse requires accurate post-processing methods. The idea behind any PNR post-processing method is to identify the basic features (such as the maximum value of the signal and the pulse area) and apply a classification algorithm. Linear classification methods (often sufficient for a small number of photons), such as principal component analysis, perform relatively poorly compared to non-linear machine learning techniques \cite{dalbec-constant_accurate_2025}. This highlights a trade-off between the algorithm's complexity and robustness. \textcite{eaton_resolution_2023} have demonstrated a resolution of $\sim$ 35 photons per single-pixel TES. The operating temperature of these detectors is usually below 100 mK to maintain low thermal noise and good energy resolution, which is especially important for long-wavelength detection and photon number resolution. \textcite{kozorezov_resolution_2006} identified a downconversion phonon noise that deteriorates the energy resolution, especially for short wavelengths.

\paragraph{\textbf{Recovery Time:}}
The recovery time of a TES detector is usually a few microseconds, limiting the pulse frequency to about the order of 1 MHz \cite{lita_counting_2008}. Lower electronic specific heat coefficient of the superconducting material, higher electron-phonon coupling, higher $T_c$, and smaller detector size contribute to faster recovery \cite{irwin_application_1995,portesi_fabrication_2015}. The downside of increasing $T_c$ is the deteriorated energy resolution. However, increasing the electron-phonon coupling can give a faster recovery without a significant change in the energy resolution \cite{calkins_faster_2011}.

\paragraph{\textbf{Timing Jitter:}}
The timing performance of the TES detector is limited by its jitter. The timing jitter depends on two components: intrinsic detector noise and external readout circuit noise. The intrinsic TES noise sets a fundamental limit on the device's timing performance, which is primarily governed by the Johnson noise of the detector and thermodynamic fluctuation noise. Additional noise sources that are non-Markovian include contact resistance fluctuation, temperature bath fluctuation, order parameter fluctuation, non-equilibrium effects, SQUID readout circuit noise, etc \cite{ascheron_transition-edge_2005}. The timing jitter values of TES detectors optimized for efficiency and PNR are usually in the 100-nanosecond range. More recently,  nanosecond range jitter values have been shown by using low input inductance SQUID amplifiers \cite{lamas-linares_nanosecond-scale_2013}.

\subsubsection{Challenges and Outlook}
An ideal TES detector design requires a low specific heat of the film to maximize the temperature change for a given amount of heat from the incident photon. The thermal conductance between the film and the substrate, as well as between the film and the electrode (electron-phonon coupling between the film and the substrate and between the film and the electrode), needs to be small enough to maintain the film temperature precisely at $T_c$ but also large enough to ensure a fast recovery time. \new{The weak thermal link between the Si cryogenic bath and the film needs to be carefully engineered, depending on the application, to maintain the right film temperature through NETF. Multiple space-efficient and high-yield methods have been demonstrated in this regard for large pixel array units\cite{ridder_fabrication_2016,bandler_lynx_2019}. However, all these methods suffer from the problem of large thermal pixel crosstalk to date. One demonstrated workaround for this problem (in the Athena/X-IFU unit design) is to have a metal coating (Au or Cu) on the side-walls of the Si grid structure\cite{miniussi_thermal_2020}. Moreover, the experimental electrothermal response of the TES detectors can not be fully explained using the simplistic model provided by equations 7 and 8: the presence of stray thermal capacitances originating from the absorber or any additional layer can drastically change the response by creating additional thermal load. To this end, one can expand the model to a two or three-body block model\cite{maasilta_complex_2012,goldie_thermal_2009} at the cost of losing a clear physical interpretation of the modelling results. The electrical interfacing of the superconducting leads to the TES film gets very difficult for densely-packed arrays. This is circumvented by stacking the contact lines vertically, spaced by thin dielectric layers\cite{yohannes_planarized_2015,devasia_fabrication_2019}. One needs to account for the stray inductances while maximizing the coupling inductance to the SQUID coil. Apart from temperature fluctuations,} TES detectors are also very sensitive to stray magnetic fields and often require proper shielding using a metal or a superconductor, which complicates the engineering of the design. TES setups optimized for efficiency can have degraded timing performance, and the ones optimized for energy resolution and low excess noise require extremely low temperatures. Also, the complex and sensitive fabrication process involved presents non-trivial issues such as device uniformity on large pixel arrays, interface purity, uniformity of the superconducting bilayer, low-loss integration of the absorber, etc. Not to mention, the SQUID readout is complex to implement and needs a cryogenic environment. Further advancements are needed to consider TES as a widespread option in the growing market for single-photon detectors. 

\subsection{Kinetic Inductance Detectors} \label{sec:KID}
\begin{figure}[h]
\includegraphics[width=2.5\linewidth]{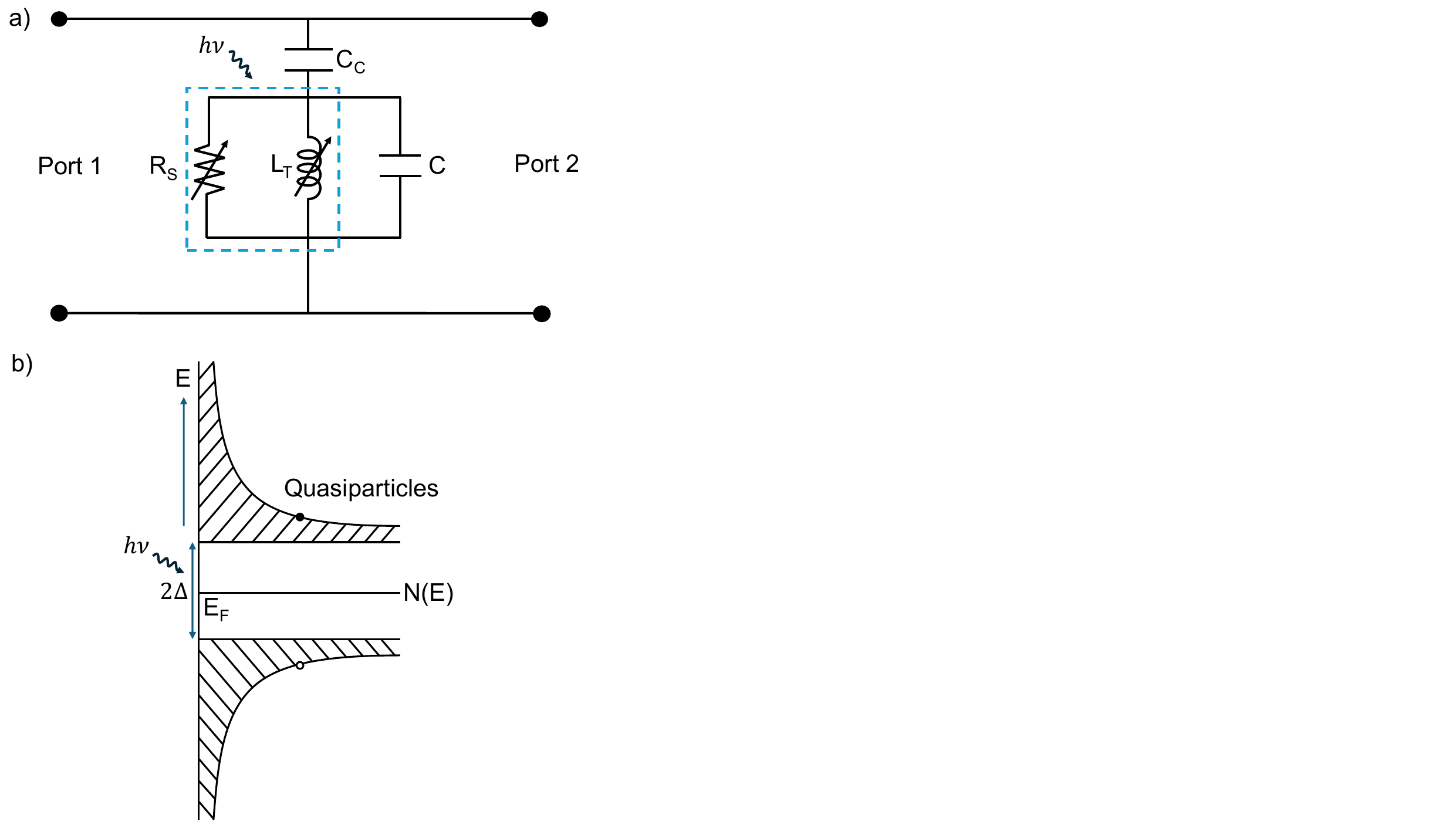}
\caption{a) Lumped element schematic of a single-pixel KID detector. \new{The blue dashed box represents the superconducting KID strip in the two-fluid model: resistance $R_s$ in parallel to the total inductance $L_T$ (sum of geometric $L_g$ and kinetic inductances $L_k$). The KID strip is patterned in parallel to a capacitance C. The total RLC circuit of the detector is coupled to the microwave feed lines using a coupling capacitance $C_c$. When a photon of energy $h\nu$ strikes the detector strip, it changes the quasiparticle density in the strip, and hence changes $R_s$ and $L_k$, which drives the detector out of resonance, giving a signal as a sharp change in the amplitude and phase of the supplied microwave.} b) Semiconductor model of the superconductor depicting quasiparticle excitations.}
\label{fig:KID}
\end{figure}

Kinetic Inductance Detectors (KIDs), also known as Microwave Kinetic Inductance Detectors (MKIDs), detect single photons by monitoring the change in surface impedance $(Z_S = R_S + i\omega L_T)$ of a thin superconducting strip \cite{day_broadband_2003}. The superconductor is patterned in parallel to a capacitor to make an RLC element that is operated under resonance by embedding it into a microwave transmission line, called the feed line. A lumped element schematic of a single-pixel KID detector is shown in Figure \ref{fig:KID}a. These devices are an attractive option for creating large-scale detector arrays because of their inherent capacity for microwave frequency division multiplexing.

\subsubsection{Detection Mechanism in KID}
When a photon is incident on the superconducting film, the energy is absorbed, and quasiparticle excitations are created (see Figure \ref{fig:KID}b). The creation of quasiparticles changes the resistance and the kinetic inductance, driving the system out of resonance. The variation in the surface impedance can be monitored through the change in amplitude or phase of the single-tone microwave probe signal centered at the resonant frequency.

\subsubsection{Performance of KID}
The system detection efficiency of KIDs directly relates to the quasiparticle generation efficiency of the incident photon. This is because the change in the number of quasiparticles dictates the change in kinetic inductance that is read out as the frequency or phase shift \cite{kutsuma2019measurement}. Using standard techniques, KIDs have been designed to maximize this energy conversion, demonstrating close-to-unity photon absorption efficiency \cite{mai2023high}. They can operate at a wide range of frequencies like any other superconducting detector; only limited by the superconducting energy gap \cite{day202425,fruitwala2020second,ulbricht2021applications,mazin2012superconducting}. Another interesting aspect of KIDs is their excellent PNR capability. Recently, Al-based KIDs were demonstrated to resolve up to 30 photons \cite{dai2025photon}. One of the main parameters that limit PNR is the energy resolution $(\Delta E)$ of the detector at a particular wavelength. $\Delta E$ depends on multiple factors such as the Fano effect, Q-factor of the resonator, amplifier noise, inhomogeneities in the detector current, readout sampling frequency, etc \cite{de2023limitations}. Additionally, a study shows that the loss of hot phonons during the initial stage of the photon downconversion process results in the degradation of $\Delta E$ \cite{de2021phonon}.

The timing performance of a KID mainly depends on the quasiparticle lifetime and noise. Particularly, the detector's recovery time depends on the quasiparticle decay time, $\tau_{qp}$, which can be estimated from the pulse width \cite{day202425}. The best reports on pulse decay time show numbers on the order of $1\mu s$. However, it is important to note that the recovery time can depend (indirectly) on the detection wavelength. For example, Al-based KID setup optimized for sensitivity at 25 $\mu m$ wavelength has a deteriorated $\tau_{qp}$ \cite{day202425}. This highlights an important trade-off between the signal-to-noise ratio (SNR) and speed: highly sensitive KIDs require longer $\tau_{qp}$ as the signal strength ($\propto \tau_{qp}$) depends on the integration time. Increasing $\tau_{qp}$ to improve SNR, however, increases the noise, increasing the timing jitter\cite{mohammad2024characterization}. The noise associated with the quasiparticle lifetime is called the generation-recombination noise ($\propto \sqrt{\tau_{qp}}$), which sets a fundamental limit on the system noise. In addition, excess phase noise, such as two-level system (TLS) noise and high electron mobility transistor (HEMT) amplifier noise, is also present, which degrades the sensitivity and timing jitter \cite{gao2007noise,gao2008physics,pappas2011two}.

\subsubsection{Challenges and Outlook}
One of the main challenges of KIDs is to overcome the noise that limits the sensitivity. TLS noise and readout amplifier noise degrade the SNR, and it remains a challenge to get close to the fundamental noise limit. TLS noise and generation-recombination noise decrease with a reduction in the temperature of operation. This trade-off between temperature and noise prompts researchers to operate these detectors at temperatures much lower than $T_c$, which makes it another issue to be considered. The trade-off between sensitivity and speed makes KIDs, currently, unsuitable for designing detector pixels that are both fast and sensitive, especially at millimeter wave radiation. Another challenge is to fabricate KID pixels with uniformity, especially in large arrays. Non-uniformity in fabrication processes can lead to faults such as $T_c$ variation, surface impedance variation owing to thickness variation, etc. \new{Creating large arrays (tens to hundreds of kilopixels) of MKIDs will require several thousands of resonator feedlines to address the detectors. This problem is solved by utilizing frequency division multiplexing, where a single microwave feedline having a uniform frequency comb can address all the detector pixels. This will reduce the enormous thermal load on the cold stage while also making the interfacing much cleaner\cite{smith2024mkidgen3}.} Complex readout electronics, pixel crosstalk, and resonator non-linearity present some of the challenges on the readout aspect of the detector. Environmental sensitivity, such as cosmic rays and stray light, also presents another hurdle to these highly sensitive detectors.
\section{Benchmarking Different SPD Platforms}\label{sec:bm}
Benchmarking single-photon detector technologies requires balancing various metrics such as dark count rate, detection efficiency, jitter, PNR capability, and operating conditions. Table \ref{tab:spd_benchmarking} summarizes the key performance metrics and trade-offs across various SPD platforms. \new{Several conclusions can be drawn from the benchmarking table regarding performance comparison of different platforms.} Commercial Silicon and InGaAs SPADs provide a mature and well-characterized performance, making them a common baseline for comparison.

Layered material-based SPDs, as well as semiconducting quantum dot and nanowire-based SPDs, remain largely in the proof-of-principle stage, with demonstration often limited to operation under specialized laboratory conditions. \new{The demonstrations to date offer a relatively low detection efficiency and slow response, compared with the commercial benchmarks. Also, apart from couple of reports, the operating temperature of these demonstrations is limited to cryogenic temperatures.} However, their tunable band structure, strong light-matter coupling, potential for room-temperature operation, \new{possible PNR capability} and heterogeneous integration capabilities, along with advances in fabrication processes, suggest pathways for improvement, even if their current figures of merit fall short of established SPD standards. 

On the other hand, SNSPDs outperform other forms of "click/no-click" SPDs, including SPADs, in several key performance metrics such as dark count rate, detection efficiency, dead time, and timing jitter. While SNSPDs do require cryogenic temperatures to operate, recent works on high-$T_c$ SNSPDs aim to open up pathways to push the operating temperature range achievable by  liquid nitrogen. 

\new{While PNR capability remains an important challenge in SNSPD, some reports achieve this capability through waveform analysis and multiplexing techniques. On the other hand, other superconducting SPDs, such as TES and KID, are the front-runners for PNR capability, maintaining excellent detection efficiency and a low dark count rate.}
However, this is achieved at the cost of operating at cryogenic temperatures. For example, the operating temperatures of TES and KID are in the several tens of mK, which allows them to have some of the best energy resolutions while maintaining the SNR and low dark counts.  These technologies have reached a maturity level where large pixel arrays can be manufactured and deployed for real-world applications.

\begin{table*}
\caption{Summary of key performance metrics of representative single-photon detectors across different platforms.\new{The
corresponding discussion sections are indicated in square bracket in the first column.}
}
\label{tab:spd_benchmarking}
\centering
\begin{ruledtabular}
\begin{tabular}{c c c c c c c c c c}
\thead{SPD type [Section]} & \thead{Material} & \thead{Wavelength \\ range \\\ [nm]} & \thead{Operating\\ Temperature \\\ [K]} & \thead{EQE \footnotemark[1]\\ (IQE \footnotemark[2]) \\\ [\%]} & \thead{Jitter \\\ [ps]} & \thead{DCR \footnotemark[3]\\\ [Hz]} & \thead{MCR \footnotemark[4]/ \\ Dead time}  & \thead{PNR \footnotemark[5]}\\
\hline
SPAD \cite{PicoQuant_PDM_Series, krahl2005performance} \footnotemark[6] & Si  & 375 - 1000 & $\sim$ 280 & 49\footnotemark[7] & $<$ 50 & $<$ 25 & 77 ns & No &  \\
SPAD \cite{Excelitas_SPCM_AQRH} \footnotemark[6]& Si  & 400 - 1060 & 278 - 343 & 70\footnotemark[8] & 350 & $<$ 1500 & 20 - 40 ns & No &  \\
SPAD\cite{IDQuantique_ID230} \footnotemark[6]& InGaAs/InP  & 900 - 1700 & 183 & 25\footnotemark[9] & 150 & 80\footnotemark[10], 200\footnotemark[11] & 2 - 100 $\mu$s & No &  \\
SPAD \cite{na2024room} \footnotemark[6]& Ge-on-Si  & 1310 & 300  & 12 & 188 & $1.5\times10^{6}$ & $-$ & No &  \\
SET \cite{komiyama2000single} [\ref{sec:qd}]& GaAs/AlGaAs  & $2\times10^5$ & 0.4  & 1 &$-$& $10^{-3}$ & $10^{3}$ Hz & No &  \\
QDOGFET \cite{rowe2006single} [\ref{sec:qd}] & InGaAs & 805 & 4  & 2 - 3 (68) &$-$& $3\times10^{-3}$ & $0.5$ Hz& No &  \\
QDOGFET \cite{shields2000detection} [\ref{sec:qd}] & InGaAs & 950  & 4  &$-$&$-$& $-$ & $0.1$ Hz & No &  \\
QDOGFET \cite{gansen2007photon} [\ref{sec:qd}] & InGaAs & 805 & 4  & 2 - 3 (68) &$-$& $3\times10^{-3}$ & $0.5$ Hz& Yes (2) &  \\
QDOGFET \cite{kardynal2007photon} [\ref{sec:qd}] & InGaAs & 684  & 4.2  & 1.3 &$-$& $10^{-4}$ & $1$ Hz& Yes (3) &  \\
QDOGFET \cite{kardynal2004low} [\ref{sec:qd}]& InGaAs & 684 & 4.2  & 0.14 (10) &$-$& $10$ & - & No &  \\
Internal defects \cite{kosaka2002photoconductance} [\ref{sec:qd}] & GaAs/AlGaAs & 700  & 4.2  & 30 &$-$& $-$ & $5\times10^{-2}$  Hz & No &  \\
RTD \cite{weng2015quantum} [\ref{sec:qd}] & GaAs/AlGaAs & 470  & 77  & 0.8 &$-$& $10^{-2}$ & $1$ Hz& Yes (2) &  \\
RTD \cite{li2007quantum} [\ref{sec:qd}] & GaAs/AlGaAs & 1310  & 4.2  & 0.23 (6.3) &$-$& $10^{-1}$ & - & No &  \\
NW \cite{luo2018room} [\ref{sec:qd}] & CdS & 457  & RT & 23 &$-$& $10^{-3}$ & $0.1$ Hz & Yes (3) &  \\
SNSPD \cite{metuh2025toward} [\ref{sec:lm}]& NbSe$_2$ & 650 - 1550 & 4 & $\sim$ 0.01 & 1100 &$-$& 135 ns	& No & \\

SNSPD \cite{zugliani2025single} [\ref{sec:lm}] & NbSe$_2$ & NIR & $\sim$ 4 &$-$& 45 &$<$ 1&$-$& No & \\

Calorimeter \cite{huang2026thermal} [\ref{sec:lm}]& Graphene & NIR / telecom & 1.2 & (87) &$-$& $<$ 1 &$-$& No \\

Calorimeter \cite{di2024infrared} [\ref{sec:lm}]& MATBG & 1550 & few K &$-$&$-$&$-$&$-$& No \\

NDC device \cite{nowakowski2025single} [\ref{sec:lm}]& BLG/hBN moiré & 675 - 11300 & 25 &$-$&$-$&$-$&$-$& No \\

Photogating \cite{roy2018number} [\ref{sec:lm}]& MoS$_2$/BLG & Visible & 80 &$-$&$-$& 0.07 &$-$& Yes \\

Photogating \cite{abraham2024room} [\ref{sec:lm}]& WSe$_2$/MoS$_2$/BP & 1550 & 300 & 21.4 &$-$& $\sim$ 720 &$-$& Possible \\

SNSPD \cite{zhang2022nbn} [\ref{sec:snspd}]& NbN &  1550  & $-$ & 90.5  & 14.7 & 50 & 21.2 ns & No &\\
 
 SNSPD \cite{marsili2013detecting} [\ref{sec:snspd}]& WSi & 1550 & 0.120  & 93 & 150 &$<$ 1 & 40 ns &No& \\
 
 SNSPD \cite{zhang2017nbn} [\ref{sec:snspd}]& NbN & 1550 & 1.8 - 2.1  & 90 - 92 & 79 & 10 & 48.5 ns &No& \\
 
 SNSPD \cite{zadeh2017single} [\ref{sec:snspd}]&  NbTiN& 1310 & 2.5  & 92 & 14.8 & $<$ 130 & 20.31 ns &No&\\

 SNSPD \cite{reddy2019exceeding} [\ref{sec:snspd}]& MoSi & 1520 - 1550  & 0.700  & 95 &$-$&$-$ & 200 ns &No&\\
 
 SNSPD \cite{reddy2020superconducting} [\ref{sec:snspd}]& MoSi & 1550 & 0.700  & 98 &$-$&$-$&$-$&No&\\
 
 SNSPD \cite{hu2020detecting} [\ref{sec:snspd}]& NbN & 1530 - 1630 & 0.8 - 2.1  & 95 - 98 &  65.8 - 106 & 100 & 42 ns &No& \\
 
 SNSPD \cite{gourgues2019superconducting} [\ref{sec:snspd}]& NbTiN & 400 - 1550 & 2.5 - 6.2  & 15 - 82 & 30 - 70 &$<$ 10 &$-$&No& \\
 
 SNSPD\cite{chang2021detecting} [\ref{sec:snspd}]& NbTiN & 1290 - 1500 & 2.5 - 2.8  & 94 - 99.5 &  15.1 & 300 - 500 & 97 ns &Possible& \\

 SNSPD\cite{cheng2023100}  [\ref{sec:snspd}]& NbN & 1550 & 1.7 & 95.5 & 50 - 90 & few-hertz & $-$ & 100\\

 SNSPD\cite{los2024high} [\ref{sec:snspd}]& NbTiN & 850 - 950 & $-$ & 94 & 7 - 11 & 40 & $-$ & Yes (7)\\
 
 SNSPD\cite{merino2023two} [\ref{sec:snspd}]& BSCCO & $\sim$ 1550 & 20 & 0.01 & $-$ & $-$ & $-$ & No\\
 
SNSPD\cite{charaev2023single} [\ref{sec:snspd}]& \makecell{BSCCO,\\LSCO-LCO} & $\sim$ 1550 & 25\footnotemark[12], 8\footnotemark[13] & $-$ & $-$ & $-$ & $-$ & No\\

 TES\cite{fukuda_titanium-based_2011} [\ref{sec:tes}]& Ti &	850 & < 0.3 & 98 (99.5) &$-$&$-$&$-$& Yes\\

 TES\cite{lita_counting_2008} [\ref{sec:tes}]& W & 1550 & < 0.1 ($T_c = 0.178$) & 95 (98) &$-$&$-$&$-$& Yes\\

 TES\cite{eaton_resolution_2023} [\ref{sec:tes}]& W & 1064 & 0.1 & > 90 & $-$ & $-$ & $-$ & Yes (37)\\

 TES\cite{calkins_faster_2011} [\ref{sec:tes}]& W & 1550 & 0.08 ($T_c = 0.15$) & $-$ & $-$ & $-$ & $\sim 460$ ns & $-$ \\

 TES\cite{lamas-linares_nanosecond-scale_2013} [\ref{sec:tes}]& W &	775, 1550 & 0.15 &$-$& 2300\footnotemark[14], 4000\footnotemark[9] &$-$&$-$& Yes\\ 

 TES\cite{fukuda_titanium_2011} [\ref{sec:tes}]& Ti & 850 & 0.1 ($T_c = 0.27$) & 98 & 25000 & 0.6 & $10^6$ Hz & Yes (8)\\

 KID\cite{mai2023high} [\ref{sec:KID}]& Al & 1518 & 0.1 & (98.9) &$-$&$-$&$-$& Yes\\

 KID\cite{dai2025photon} [\ref{sec:KID}]& Al & 1550, 1064 & 0.05 & 0.037\footnotemark[9], 0.056\footnotemark[15] &$-$&$-$& 56 ns & \makecell{Yes (32\footnotemark[9], \\25\footnotemark[15])}\\

\end{tabular}
\end{ruledtabular}
\footnotetext[1]{External Quantum Efficiency}
\footnotetext[2]{Internal Quantum Efficiency}
\footnotetext[3]{Dark Count Rate}
\footnotetext[4]{Maximum Count Rate}
\footnotetext[5]{Photon Number Resolution}
\footnotetext[6]{Commercially available SPDs}
\footnotetext[7]{at 550 nm}
\footnotetext[8]{at 650 nm}
\footnotetext[9]{at 1550 nm}
\footnotetext[10]{for 10\% efficiency}
\footnotetext[11]{for 20\% efficiency}
\footnotetext[12]{BSCCO}
\footnotetext[13]{LSCO-LCO}
\footnotetext[14]{at 775 nm}
\footnotetext[15]{at 1064 nm}
\end{table*}

\section{Applications of SPD}\label{sec:app}

The ability of single photon detectors to detect very weak light signals down to single photons and count the photon numbers has found extensive application in cutting edge science and technology. 
While the main thrust in research to develop next generation technology applications with single photon detectors has been in quantum communications and computing, these detectors have found applications in many other fields like fundamental scientific research, medical and industrial imaging, and sensing and ranging.

\subsection{Quantum Communication and Computation}

\paragraph*{Quantum Communication:} Quantum communication refers to transmission and receiving of quantum information in the form of qubits \cite{couteau2023applications}.
Photons are suitable to encode the information in the form of qubits which can be used to communicate information between two points in space. In this domain, high quality single photon sources as well as detectors play a very important role \cite{nielsen2004optical}.
Quantum cryptography is a subdomain in quantum communications where single photon detectors and sources are used to secure classical data sources from manipulation and eavesdropping.

The broader area of quantum communications refers to a \emph{quantum internet} where various quantum devices communicate with each other over a large mesh of interconnected devices. The lossless transmission and detection of single photons is very crucial for this application. To achieve this goal in existing deployments of single-mode fiber networks, the photons used for communication must be in S band, C band, and L band in the range of 1460 nm and 1625 nm, with C band centered around 1550 nm, since the loss in optical fibers is minimum in these bands \cite{johnson2009optical}. Several single photon detectors targeting these bands have been demonstrated \cite{komiyama2000single,abraham2024room}. On the other hand, optimum wavelength for free space communication (for example, satellite communication) is around 800 nm \cite{bourgoin2013comprehensive}, which can be covered by both silicon SPAD as well as SNSPDs. Room temperature detectors have deployment edge  at the end user side \cite{abraham2024room}. High timing precision (< 200 ps) is also very important for high-speed links \cite{yuan2008gigahertz}. Although III-V technology-based avalanche photodetectors were conventionally used for quantum communication \cite{zhang2015advances,jiang2014inp}, this scenario is fast emerging with superconducting single photon detectors replacing the conventional technology \cite{slysz2007fibre,verevkin2004ultrafast,bhaskar2020experimental,you2020superconducting,ruskuc2025multiplexed}.

\paragraph*{Quantum Cryptography:} Quantum cryptography refers to use of quantum laws to encrypt data. This technology can be used to encrypt both classical as well as quantum information. The main advantage in this method lies in the fact that a quantum state changes on measurement. Hence, any attempt of eavesdropping can be detected, since the quantum state of transmitted photon changes \cite{ekert1991quantum}. Quantum cryptography has, to date, been the most technologically advanced domain in the field of quantum communications \cite{couteau2023applications}. Quantum key distribution (QKD) refers to distribution of encryption keys in a communication channel using laws of quantum physics. Many algorithms have been developed to achieve the same, an important one being the Ekert protocol \cite{ekert1991quantum}. \new{Low timing jitter, high count rate and low DCR are the key performance metrics for SPDs for fast and reliable quantum communication system implementation.} Although QKD has been the most secure solution for data encryption, a major limitation has been the lower data rates \cite{grunenfelder2023fast}. A lot of research is ongoing to improve data rates and SNSPD detectors are being studied to achieve the desired data rates because of their higher timing precision \cite{grunenfelder2023fast,beutel2021detector,sax2023high}. 
Superconductor based detectors have been at the forefront of the research for single photon detectors for next generation quantum cryptography \cite{takesue2007quantum,takemoto2015quantum,hadfield2006quantum,shibata2014quantum}. 

\paragraph*{Quantum Computing:} Quantum computing is another fast-emerging area of application of single photon detectors. Quantum computing refers to implementing useful computational algorithms using quantum technologies instead of the more traditional von-Neuman or the neuromorphic computer architectures. Single photons play a crucial role in this domain as they are used to implement quantum gates \cite{couteau2023applications}. The efficient control of quantum states of single photons represented by polarization, energy and other degrees of freedom of the photon is important for quantum computing. The three main components of a quantum computing architecture \cite{na2024room} are (1) single photon sources, (2) quantum gates and circuits and (3) single photon detectors. SNSPDs, have been at the forefront of the detector technologies used in quantum computing platforms because of their superior performance metric \cite{na2024room}. In particular, NbN nanowires have been the most used material in single photon detectors for quantum computing platforms \cite{qin2025integrated}. The major challenge in single photon detectors for quantum computing platform remains is their requirement to operate at higher temperatures and frequencies. 
\new{Low DCR and high detection probability are required to reduce errors in quantum computing systems. In addition, high photon count rate helps improve the overall computing speed.}
Due to maturity in technology, silicon based or III-V based systems have been preferred for quantum computing \cite{rudolph2017optimistic,baldazzi2025four,gonzalez2021scaling,langford2011efficient}. 
However due to higher efficiency of SNSPDs for single photon detectors, a mixed approach has been recently demonstrated \cite{psiquantum2025manufacturable} where NbN SNSPDs are integrated into the silicon based quantum computing platform, where SNSPDs are used for photon detection, while other photonic components like waveguides, couplers, etc., utilize the more mature silicon photonic technologies.

\subsection{LIDAR and Remote Sensing}

The Light Detection and Ranging (LIDAR) uses the time of flight measurement of reflected optical signal to measure the distance from an object \cite{collis1970lidar}. 
Owing to the shorter wavelengths of the LIDAR systems are in general more capable of higher resolution imaging of the objects at longer distances compared to their radio frequency counterpart: RADAR (Radio Detection and Ranging) \cite{mcmanamon2016laser}. 
The LIDAR systems have found extensive applications in airborne platforms as well as autonomous vehicle navigation systems \cite{itzler2017geiger}.

\new{Single photon detectors have been used to significantly improve the performance of LIDAR systems as they have high sensitivity and long distance range \cite{qian2023single}.}Time-correlated single-photon counting (TCSPC)  has been used in LIDAR systems owing to the high sensitivity and temporal resolution provided by this technology \cite{massa1998time,tobin2021robust}. 
Single photon detectors with high efficiency and low jitter are crucial in such systems. \new{Low DCR is crucial for LIDAR systems as the background illumination is usually very weak. Since remote sensing systems are deployed in varied environments, stability to factors such as temperature is also important.}
Due to their high efficiency and low jitter, SPADs and SNSPDs have been the most used detectors in this domain \cite{vodolazov2019minimal,tobin2021robust,mccarthy2013kilometer}.Many interesting applications have been demonstrated using single photon LIDARs. These include, but not limited to, imaging  dense forest terrains \cite{swatantran2016rapid}, 3D imaging upto 200 km \cite{li2021single} and real time 3D imaging of moving scenes in scatterer obscured environment \cite{tobin2021robust}.

\subsection{Imaging and Spectroscopy}
Biomedical medical imaging and optical spectroscopy methods have played a very significant role in clinical diagnosis and therapy \cite{marcu2012fluorescence}. Among numerous imaging technologies, fluorescence-lifetime imaging microscopy (FLIM) is a crucial spectroscopic and imaging method for studying the various lifetimes of light emitted by fluorophores present in tissue proteins of the human body. This information is then analysed to identify normal tissues from diseased tissues \cite{marcu2012fluorescence,marcu2014fluorescence}. Changes in spectral properties of emitted light have been used to diagnose osteoarthritis, liver fibrosis, cardiovascular conditions and many forms of cancer \cite{lagarto2020real}. FLIM in general uses the TCSPC method to measure the decay rate of photon emission from the fluorophores in the sample \cite{7407375,marcu2014fluorescence,lagarto2020real}. To this end, SPDs with high quantum efficiency and low timing jitter \cite{korzh2020demonstration} can provide picosecond timing accuracy, making the acquired data highly reliable and accurate \cite{lagarto2020real,meleshina2016probing}, which meets the state-of-the-art requirements of modern clinical diagnosis. \new{Further, large-scale SPD arrays\cite{zickus2020fluorescence} can advance the capabilities of FLIM by wide-field and high-speed acquisition at higher spatial resolutions to capture intricate multimodal details simultaneously for monitoring dynamic cellular processes and in vivo diagnostics. In super-resolution imaging techniques which minimize fluorescence photons needed for high localization precision\cite{balzarotti2017nanometer}, high sensitivity of SPDs can improve spatial resolution. For example, in MINFLUX method\cite{balzarotti2017nanometer}, the photons emitted from a flourophore upon excitation from a doughnut shaped beam with a local minimum are counted at various beam positions and the exact location of flourophore is reconstructed within a precision of 1 nm, while ideally detecting a single photon from the excitation minimum is enough to determine the coordinate.}

Photon Counting Computed Tomography (PC-CT) is another important biomedical imaging technique where single photon detectors find their application \cite{flohr2020photon,douek2023clinical}. Photon-counting detectors are key to this technology, where single photons of X-ray radiation are typically detected by high atomic number semiconductors, such as cadmium telluride (CdTe) or cadmium zinc telluride (CZT) \cite{flohr2020photon}.
PC-CT technology offers several advantages over conventional computed tomography (CT), including reduced noise and improved spatial resolution \cite{douek2023clinical}. 
The most distinct advantage of PC-CT is, however, its capability to resolve the energy of the incoming X-ray photons, which helps in generating a \emph{colour} image of the sample compared to the \emph{grey-scale} image generated by conventional CT \cite{douek2023clinical,cormode2017multicolor}. PC-CT, with its spectral resolution capabilities, has been used to image various human organs to extract useful information for clinical diagnosis \cite{poon2022first,si2025lung,boccalini2025spectral,si2019spectral}.

 Astronomical imaging also relies heavily on single-photon detectors and counters for imaging from X-rays to the far-infrared. Imaging increasingly distant and faint celestial objects, such as galaxies and pulsars, requires highly sensitive and ultra-low noise detectors. \new{Improving SDE and reducing DCR of the SPDs are crucial for collecting the signal from an ultra-low light setting while maintaining reasonable SNR.} X-ray Integral Field Unit of Advanced Telescope for High-Energy Astrophysics (Athena) mission uses Mo/Au TES with Au and Bi absorbers for detecting faint extended sources with as low instrumental noise as possible \cite{pajot2018athena,barret2023athena}. Dark-speckle Near-IR Energy-resolved Superconducting Spectrophotometer (DARKNESS)\cite{meeker2018darkness} and the MKID Exoplanet Camera (MEC)\cite{walter2020mkid} use large arrays of MKIDs for high contrast spectroscopy of exoplanets' spectra in the 800-1400 nm band. Infrared telescopes such as the James Webb Space Telescope (JWST)\cite{gardner2006james,rieke2005overview} use semiconductor-based detectors (mercury-cadmium-telluride and arsenic-doped silicon detectors used in JWST) for near-IR and mid-IR wavelength ranges. Background Imaging of Cosmic Extragalactic Polarization (BICEP)\cite{lange2003background}, the Cosmology Large Angular Scale Surveyor (CLASS)\cite{essinger2014class}, and the South Pole Telescope (SPT)\cite{ruhl2004south} are some facilities that utilize TES arrays to image the cosmic microwave background radiation. SNSPDs are also widely used for various deep-space applications. One such application is NASA’s Deep Space Optical Communications project (DSOC) \cite{biswas2025overview}, which utilizes free-space-coupled SNSPDs to enable communication between spacecraft and ground-based control centers. SPADs and EMCCDs are very useful in instruments, such as ULTRASPEC\cite{ives2008ultraspec} and Aqueye+\cite{zampieri2015aqueye+}, designed for fast imaging/spectroscopic applications.

Single-photon detectors play a central role in a wide range of spectroscopic and scientific laboratory techniques, where the detection of weak optical signals down to the single-photon level with high spectral and temporal resolution is crucial \cite{becker2005advanced}. Time-resolved photoluminescence (TRPL) is one such example where statistics are built using the time difference between the synchronization pulse from a pulsed laser and the photon emitted from the sample. This allows one to perform lifetime measurements down to several ps, and is usually limited by the jitter of the photodetector \cite{stevens2006time} and the TCSPC electronics \cite{becker2005advanced}. SPDs are also highly popular in performing single-photon-based quantum optics experiments, such as $g^{(1)}$ correlation for coherence measurement \cite{migdall2013single}, $g^{(2)}$ correlation for testing the purity of single-photon sources \cite{migdall2013single}, and Hong-Ou-Mandel measurements for gaining knowledge about the indistinguishability of photons \cite{gerrits2015spectral}. Arrays of SPDs are becoming increasingly popular for various ultra-low signal optical spectroscopy \cite{tye2025time} and metrology \cite{mccarthy2025high}techniques. 

\section{Conclusion}\label{sec:con}
In summary, while conventional SPADs remain a mature technology with room temperature operation, low-dimensional platforms offer pathways to innovate and improve SPDs in other aspects of performance for new frontiers of quantum technologies and advanced imaging applications. They offer the unique advantages of quantum confinement, extreme tunability, ultrafast carrier dynamics, phase change, and seamless integration with photonic circuits that conventional platforms cannot match. 

As seen from the performance benchmarking section, SNSPDs clearly outperform in the infrared regime with a higher detection efficiency, lower timing jitter, dead time and dark count rate, besides the scope for large scale active detection area. Due to this attractive potential of SNSPDs, some of the high-priority directions are geared towards engineering multipixel layouts, readout and multiplexing schemes with appropriate thermal management. For upscaling, pursuits on understanding microscopic detection mechanisms (non equilibrium quasiparticles, phase slips, vortex dynamics) are essential to build predictive models for device performance. To push the temperature limits, there is a need to actively explore SNSPD architectures with emerging high-Tc superconductors. Other superconducting devices based on TES and KID offer better PNR capabilities with promising scalability shown in imaging and spectroscopy applications. To overcome slow recovery times in TES, approaches toward machine learning for pile-up discrimination algorithms to resolve multiple photon pulses within a single thermal spike can help increase detection rates without complex hardware changes.  Adding PNR capability to an SNSPD is another important future research direction.

For applications such as QKD and QRNG which need efficient SPDs along with  extremely low dark count rate and minimal afterpulsing probability without cryogenic cooling, layered materials and their heterostructures are potential candidates for higher temperature operation as they offer detection mechanisms beyond superconducting paradigms. Continued efforts in heterogeneous integration, optical engineering, scaling material synthesis, optimizing device contacts, and on-chip coupling are crucial in overcoming intrinsic limits and extending the interesting physical mechanisms to reliable device architectures. 

With a holistic approach spanning advancements at the different levels of engineering - device, circuit and data processing, low-dimensional single photon detection technologies can transition from laboratory demonstrations to real-world deployable systems.

\textbf{Acknowledgements}
M.D. acknowledges startup funding support from JNCASR. K.M. acknowledges support from National Quantum Mission, an initiative of the Department of Science and Technology (DST), Government of India, a grant from Indian Institute of Science under IoE, a grant from DRDO, and a grant from I-HUB QTF, IISER Pune.
\bibliography{references}

\end{document}